\documentclass[floatfix,footinbib]{iopart}
\eqnobysec
\usepackage{graphicx}
\usepackage{iopams}
\usepackage{eqnarray}
\usepackage{mathrsfs}
\usepackage{wasysym}
\usepackage{hyperref}
\usepackage[usenames,dvipsnames]{xcolor}
\usepackage[normalem]{ulem}
\usepackage{tikz}

\renewcommand{\thesection}{\arabic{section}}

\graphicspath{{./Figs/}}

\def\Bbb{\mathbb}
\def\Cal{\mathcal}
\def\Cyc{\mathscr}
\def\Str{\mathcal}

\newcommand{\eqdef}{ {\kern 0.2em}={\kern -0.5em}:{\kern 0.2 em} }
\newcommand{\defeq}{ {\kern 0.2em}:{\kern -0.5em}={\kern 0.2 em} }

\newcommand{\expct}[1]{\langle{#1}\rangle}

\newcommand{\Real}{{\mathbb R}}

\newcommand{\Int}{{\mathbb Z}}

\newcommand{\downtri}{\raisebox{1pt}{$\bigtriangledown$}}
\newcommand{\dntri}{\downtri}
\newcommand{\uptri}{{$\bigtriangleup$}}
\newcommand{\BZ}{\mathrm{BZ}}
\newcommand{\xfer}{{\Bbb T}}

\newcommand{\FS}{\Omega}
\newcommand{\BZO}{{\mathrm BZ}\setminus 0}
\newcommand{\Nfluct}{\tilde{N}}
\newcommand{\Nfluctp}{\tilde{N}_{>}}
\newcommand{\Nfluctm}{\tilde{N}_{<}}
\newcommand{\Latt}{\Int_C}
\newcommand{\setof}[2]{\left\{{#1}\,\middle|\, {#2}\right\}}

\newcommand{\Vac}{{\varnothing}}
\newcommand{\ket}[1]{|{#1}\rangle}
\newcommand{\bra}[1]{\langle{#1}|}

\newcommand{\braket}[3]{\langle{#1} | {#2} | {#3}\rangle}
\newcommand{\partfn}{{\mathcal Z}}
\newcommand{\NN}{{\mathsf N}}
\newcommand{\Nspin}{{\NN}_s}
\newcommand{\BZave}[1]{\int_0^\pi {#1}\, \frac{dq}{\pi}}
\newcommand{\paren}[1]{\left({#1}\right)}

\newcommand{\nn}{\mathsf{n}}
\newcommand{\jj}{\mathsf{j}}
\newcommand{\dual}[1]{{}^\dagger{#1}}
\newcommand{\Sat}{\ensuremath{\mathsf{Sat}}}
\newcommand{\FM}{\ensuremath{\mathsf{FM}}}

{\color{#1}}%

\begin{document}

%%%%%%%%%%%%% string motifs and bond glyphs
\newsavebox{\emptydn}
\sbox{\emptydn}{\raisebox{-0.1em}{\includegraphics[width=12pt]{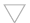}}}
\newsavebox{\upleft}
\sbox{\upleft}{\raisebox{-0.1em}{\includegraphics[width=12pt]{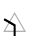}}}
\newsavebox{\upright}
\sbox{\upright}{\raisebox{-0.2em}{\includegraphics[width=12pt]{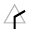}}}
\newsavebox{\downright}
\sbox{\downright}{\raisebox{-0.2em}{\includegraphics[width=12pt]{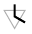}}}
\newsavebox{\downleft}
\sbox{\downleft}{\raisebox{-0.2em}{\includegraphics[width=12pt]{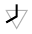}}}
\newsavebox{\creation}
\newsavebox{\annihilation}
\sbox{\annihilation}{\includegraphics[width=12pt]{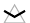}}
\sbox{\creation}{\includegraphics[width=12pt]{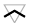}}
\newsavebox{\ulline}
\sbox{\ulline}{\raisebox{-0.1em}{\includegraphics[width=6pt]{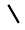}}}
\newsavebox{\urline}
\sbox{\urline}{\raisebox{-0.05em}{\includegraphics[width=6pt]{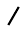}}}
\newsavebox{\horline}
\sbox{\horline}{\raisebox{0.15em}{\includegraphics[width=6pt]{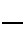}}}
%%%%%%%%%%%%%%%%%%5
%\newcommand{\eqpar}{\stackrel{2}{=}}
\newcommand{\eqpar}{\,{=}_{2}\,}
\newcommand{\mod}{\mathrm{mod}\,}
%%%%%%%%%%%%% 

\title[Triangular lattice antiferromagnetic Ising model]{Classical triangular lattice antiferromagnetic Ising model as a free-fermion/superconductor system}

\author{Amir Nourhani$^{1,2,3,4,5}$, Vincent H Crespi$^{6,7,8}$, \\ and Paul E Lammert$^{6}$}
\address{$^1$ Department of Mechanical Engineering, University of Akron, Akron, Ohio  44325, USA}
\address{$^2$ Department of Biology, University of Akron, Akron, Ohio  44325, USA}
\address{$^3$ Department of Mathematics, University of Akron, Akron, Ohio  44325, USA}
\address{$^4$ Department of Chemical, Biomolecular, and Corrosion Engineering, University of Akron, Akron, Ohio  44325, USA}
\address{$^5$ Biomimicry Research and Innovation Center, University of Akron, Akron, Ohio  44325, USA}
\address{$^6$ Department of Physics, Pennsylvania State University, University Park, Pennsylvania 16802, USA}
\address{$^7$ Department of Materials Science and Engineering, Pennsylvania State University, University Park, Pennsylvania 16802, USA}
\address{$^8$ Department of Chemistry, Pennsylvania State University, University Park, Pennsylvania 16802, USA}
\ead{nourhani@uakron.edu, lammert@psu.edu}
\begin{abstract}
We present a treatment of the triangular lattice antiferromagnetic
Ising model (TAFIM) based on a small number of
elementary ideas common to statistical and solid-state physics.
The TAFIM is represented as a reduced BCS model in one space,
one (imaginary) time dimension.
The representation is approximate for nonzero temperature, but allows
quick derivation of asymptotically exact thermodynamic functions, and
the divergence of the spin-spin correlation length.
The fermionic representation is exact at zero temperature.
We demonstrate the existence of a two-dimensional continuum of
zero-temperature equilibrium macrostates characterized by satisfied bond
fractions of the three different orientations, and calculate their entropy densities.
\end{abstract}

\date{March 28, 2023}

\maketitle 

\section{Introduction
\label{sec:intro}}

Classical Ising spin models with nearest-neighbor interactions are 
archetypes of many fundamental phenomena of statistical mechanics.
For instance, the model on a bipartite lattice with ferromagnetic couplings
exemplifies order-disorder transitions, spontaneous symmetry
breaking~\cite{Peierls-36}, and nonclassical critical phenomena~\cite{Onsager-44},
The \textit{anti}ferromagnetic model on a triangular lattice\cite{Wannier-50,Houtappel-50},
however, behaves completely differently.
Even at zero temperature, it fails to order, because there is no spin configuration
that satisfies all three bonds around even a single elementary triangle. 
As the simplest model with this property, the triangular lattice antiferromagnetic Ising
model (TAFIM) is the archetype of geometric \textit{frustration},
%(the coinage is due to P.~W.~Anderson, according to Toulouse\cite{Toulouse77}).
the presence of incompatible but equally strong elementary interactions.
Frustration occurs in an enormous range of systems, from 
water ice\cite{Pauling-35,Giauque+Stout-36} to spin systems\cite{Toulouse77,Moessner01,Normand09,Gingras14,Starykh15, Schmidt17}, artificial spin ice\cite{Wang04,Zhang+12,Perrin16}, colloidal assemblies\cite{Tierno16, Han08}, Coulomb liquids\cite{Mahmoudian15}, lattice gases\cite{Weight03}, ferroelectrics\cite{Choudhury11}, coupled lasers\cite{Nixon13}, and self-assembled lattices of microscopic chemical reactors\cite{Wang+Fraden-16}.
For this reason, frustration is of interest to the broad spectrum of condensed matter
physicists, and a treatment of
the TAFIM based on basic concepts
common to statistical and solid-state physics is desirable.
This paper aims to provide such a treatment.
The term ``frustration'' was coined by P.~W.~Anderson
according to Toulouse\cite{Toulouse77}.
The present work might be called ``poor man's TAFIM'', in homage\cite{Anderson-70}.

The TAFIM has been studied for decades. Integral expressions for exact
thermodynamic functions were determined in
1950\cite{Wannier-50,Houtappel-50,Lavis+Bell-v1}.
Wannier determined\cite{Wannier_erratum} the residual entropy.
Spin correlation functions were studied intensively by
Stephenson\cite{Stephenson-64,Stephenson-66,Stephenson-70a,Stephenson-70b}.
The representation of zero-temperature TAFIM configurations by strings
goes back at least to the equivalence with solid-on-solid models pointed out
by Bl\"{o}te and Hilhorst\cite{Blote+Hilhorst-82} in connection with
striped phases\cite{Pokrovsky+Talapov-79}, which were a focus of interest
in the early 1980's (see den Nijs \cite{den-Nijs-88} for a review).  

What is offered here against that history is: a special focus on
zero-temperature equilibrium macrostates,
bringing out an astonishing richness which
has not been explicitly discussed in the literature,
and an especially simple and physically appealing approach,
based on a widely familiar fundamental paradigm.
The cost of this simplicity is an only approximate treatment of nonzero temperature.
The approximation is nevertheless well-motivated and turns out to be surprisingly good,
even up to infinite temperature.
Using a mapping of bond configurations to trajectories of a system of fermions
in one space, one (imaginary) time dimension, bulk thermodynamic properties
(entropy and energy) are accessible through ground-state energies of
the fermionic systems, and spin-spin correlation functions through fermion
density fluctuations.

At $T=0$, the TAFIM is equivalent, in a sense,
to a simple Fermi gas. Still, it has surprising aspects.
Traditional approaches are well-suited only to taking the zero-temperature
limit \textit{after} the thermodynamic limit.
Consequently, they find only a single zero-temperature macrostate,
albeit with a nonvanishing entropy density. Our approach easily takes
the thermodynamic limit \textit{at} $T=0$, and finds a \textit{continuum}
of distinct macrostates. At low temperatures, and below some crossover
size, this can be the more accurate description.

For $T>0$, the Hamiltonian of our equivalent fermion system develops
pairing terms, and is thus a superconductor.
Schultz, Mattis and Lieb\cite{Schultz+Mattis+Lieb} observed the analogy with
BCS theory in their famous work on the square-lattice Ising model.
It is much more forceful in the present case because of both the simplicity
of the transformation to fermionic representation and the occurrence of the
normal state at zero temperature. Perhaps some non-exactly-solvable
frustrated models can be handled similarly.

We consider the essence of TAFIMs to be that every elementary triangle
is frustrated, rather than that every bond is antiferromagnetic, so
with all possible choices of anti/ferromagnetic bond patterns with that property.
Dealing with larger class of models turns out to be hardly any more work,
once one describes configurations in terms of un/satisfied bonds
rather than spins. All fully-frustrated bond patterns are locally equivalent,
but there are nonlocal distinctions. We emphasize their topological nature.
For example, the fermionic model corresponding to a given cylindrical
TAFIM supports states only of even particle number, or odd particle
number, depending upon the classification of the bond pattern.

%%%%%%%%%%%%%%%%%

\section{Preview}
%%%%%%%%%%%%%%%%%%%%%%%%%% 
\begin{figure}
\includegraphics[width=95mm]{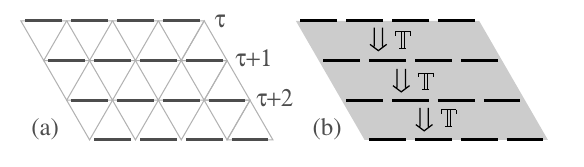}
\caption{The transfer matrix idea. In the partition function,
  the sum over interstitial (light gray) bonds in (a) reduces to
  the transfer matrix $\xfer$ acting between row configurations.
\label{fig:transfer-matrix}}  
\end{figure}
%%%%%%%%%%%%%%%%%%%%%%%%%% 

The most specialized statistical mechanics notion that
we use is that of \textit{transfer matrix}.
Although widely known nowadays, an adapted refresher may be helpful.
Suppose the bonds of a triangular lattice can be either satisfied or unsatisfied.
Orient the lattice as in Fig.~\ref{fig:transfer-matrix}(a) so that rows
of horizontal bonds labelled $\cdots,\tau,\tau+1,\cdots$ and shaded darkly
are separated by interstices of slanted bonds.
If $X$ denotes a configuration of all the bonds, the partition function is
\begin{equation}
  \nonumber
  \partfn = \sum_X^* e^{-\beta E(X)}.
\end{equation}
  The star indicates that the sum is constrained. Not every set of 
  bonds is the satisfied set for some configuration. The restriction here
  is the subject of Section~\ref{sec:fully-frustrated}.
  The idea now is to sum over configurations in a staged
way: an outer sum over the configurations $\cdots,X_\tau,X_{\tau+1},\cdots$ of the
rows of horizontal bonds, and an inner sum over the configurations on the
interstices. Since contraints on an interstitial bond depend on no
horizontal bonds except those on the bordering rows,
the partition function becomes
\begin{equation}
  \nonumber
  \partfn = \sum_{\cdots,X_\tau,X_{\tau+1},\cdots} \cdots
  \braket{X_\tau}{\xfer}{X_{\tau+1}} \braket{X_{\tau+1}}{\xfer}{X_{\tau+2}} \cdots,
\end{equation}
where $\xfer$ is the transfer matrix. It is certainly a matrix, indexed by
configurations on two neighboring rows.
Taking the Dirac notation seriously, we try to interpret the
row configuration $X_\tau$ as a state vector, as in quantum mechanics.
This turns out to be a good idea.
Another one is to identify satisfied horizontal bonds with particles,
taking them to be fermions to prevent them from sitting on the same
location. Then, if the system has $L$ rows, and we use periodic boundary conditions,
the partition function is $\Tr \xfer^L$. We complete the transition to a
quantum mechanical idiom by identifying $\xfer$ with $e^{-H} = e^{-iH(-i)}$, the
evolution operator through time $\Delta t = -i$ for an appropriate Hamiltonian
of the system of fermions living on a one-dimensional lattice.
Then, the classical partition function $\partfn$ of a two-dimensional lattice
of spins becomes a quantum mechanical partition function $\Tr e^{-HL}$ of a
system of fermions on a one-dimensional lattice at temperature $L^{-1}$.

At zero temperature, the fermions are \textit{semi}-conserved, in the sense
that they can disappear in pairs with the passage of imaginary time, but
not appear. Otherwise, the fermions are noninteracting, and appropriate
boundary conditions can suppress the annihilation process,
in which case particle number is strictly conserved.
This implies significant influence of boundary conditions on the bulk
equilibrium macrostate even in the thermodynamic limit.
On a cylinder of fixed circumference, taking the length to infinity,
we obtain distinct bulk states with distinct entropy densities, depending
on the number of fermions (See Fig.~\ref{fig:entropy-curve}).
In the planar thermodynamic limit, 
there is a two-dimensional continuum of ``equilibrium'' macrostates
(See Fig.~\ref{fig:entropy-surface})
distinguished by the fractions of frustrated bonds with each of the three orientations,
that is, not only microscopic degeneracy reflected in nonzero entropy density,
but \textit{macroscopic} degeneracy.
This depends on taking the limit $T\to 0$ before the thermodynamic limit.
If the zero temperature limit is taken after the thermodynamic limit,
only the unique one of these macrostate with maximal entropy is
accessible\cite{Aizenman+Lieb-81}, hence $T\to 0$ limits of exact thermodynamic
functions\cite{Wannier-50,Lavis+Bell-v1} are insensitive to all the other
macrostates.
We relate the entropy densities of these equilibrium macrostates to
energy densities of fermion ground states and calculate them
in the form of series expansions in Section \ref{sec:integrals}. 

Nonzero temperature (of the \textit{spin} system) is studied in Section~\ref{sec:T>0}.
In the fermionic representation, it implies having not just pair annihilation,
but pair creation ---
the TAFIM at nonzero temperature is (equivalent to) a \textit{superconductor}.
We work out a simple low-temperature approximation, which
turns out to be surprisingly accurate to high temperature.
Fig.~\ref{fig:approx-vs-exact-thermo} compares
the results of our approximate theory to the exact thermodynamic functions.
The low temperature asymptotics of the approximate theory
[Eqs. (\ref{eq:thermo-asymptotics})] are exactly right.
Even the superconducting order parameter $\expct{c_mc_{m+1}}$ is meaningful,
being directly porportional to the energy density of the TAFIM (\ref{eq:e/J}).

Zero temperature is a critical point of TAFIM systems in that the spin correlation
length diverges as $T\to 0$.
Stephenson\cite{Stephenson-64,Stephenson-70a} discovered that
the two-spin correlation function falls off only as $r^{-1/2}$ at zero temperature.
Section~\ref{sec:spin-spin-correlator} studies spin correlations 
via density fluctuations in the equivalent fermion system.
The cited $T=0$ asymptotic behavior is recovered exactly.
For $T>0$, and using the approximate theory, we
find that the correlation length diverges as $e^{2\beta J}$ as $T\to 0$.
This agrees with previous determinations\cite{Stephenson-70a,Wojtas+Millane-09}.
However, our numerical coefficient is about 20\% smaller.

\begin{figure*}
\includegraphics[width=1.2\textwidth]{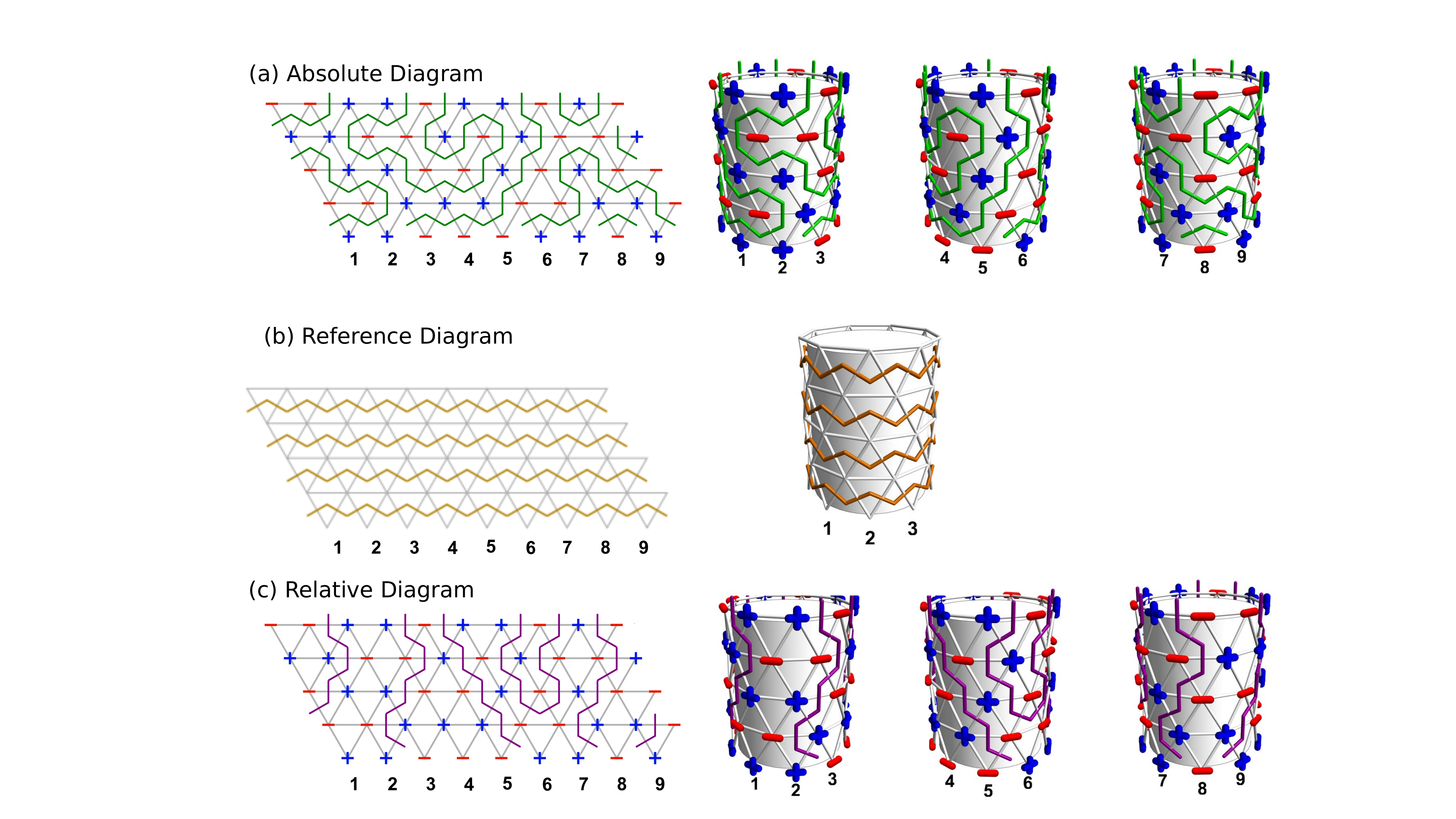}
\caption{
(a) A configurational absolute string diagram $\Cal{D}$ for a strict TAFIM.
(b) The reference diagram $\Cal{D}_{\mathrm{zigzag}}$ (see Section~\ref{sec:relative-diagrams}.
(c) Relative diagram $\Cal{D}+\Cal{D}_{\mathrm{zigzag}}$.
\label{fig:cylinders}}  
\end{figure*}

But how does the fermionic representation arise at all?
First, focus on which bonds are satisfied, rather than spins.
This is fruitful because the constraints on the set of satisfied bonds
are simple. They may be viewed as collections of strings of
dual edges with no ends inside the system (See Fig.~\ref{fig:cylinders}(a)).
In addition, there are even/odd restrictions related to the topology of
the surface on which the model is defined.
This rewriting can be carried out (Section~\ref{sec:fully-frustrated})
for arbitrary fully-frustrated triangulated surface.
Each elementary triangle is frustrated, but the system does not necessarily
look like a regular lattice, even locally.
On a cylinder, and at zero temperature, the string representation can be modified
so that the modified strings are interpreted as
particle worldlines, with the length of the cylinder corresponding to ``time''.
This is done in Section~\ref{sec:ground}.
The point of zero temperature is that the particles are essentially noninteracting.
It is then easy to write down directly (Section~\ref{sec:fermions})
a Hamiltonian that generates that allowed set of worldlines.

%%%%%%%%%%%%%%%%%%%%%%%%% 
%%%%%%%%%%%%%%%%%%%%%%%%%% 
%%%%%%%%%%%%%%%%%%%%%%%%%%
\section{From spins to strings: fully-frustrated triangulated surfaces}\label{sec:fully-frustrated}

\subsection{first ideas}

Graphical representations have a distinguished history in statistical mechanics
of spin systems.
The high-temperature expansion for a \textit{ferromagnetic} Ising model
is conveniently described by closed loops of bonds.
The standard low-temperature expansion uses the dual edges crossing
{unsatisfied} bonds (these are domain walls).
For systems with {quenched} disorder, such as spin glasses,
it can be helpful to mark {unsatisfied} bonds.
The resulting strings have ends in the frustrated plaquettes,
but this is not a useful approach for a fully-frustrated system.
This section develops such representation for arbitrary fully-frustrated
triangulated surface (e.g., a portion of the plane with holes cut in it,
a cylinder, a torus, a higher-genus or even nonorientable surface).
``Fully-frustrated'' means that each elementary triangle is frustrated,
i.e., contains one or three AFM bonds.
This generality highlights the topological aspects.

Stating the main result requires some definitions.
In each elementary triangle and bond, choose a centre.
If the surface is planar, bond $b$ corresponds to a \textit{dual edge}
$\dual{b}$.
If $b$ borders two triangles $\dual{b}$ is the line segment joining their centres
(Fig.~\ref{fig:primal-dual}).
Otherwise, $\dual{b}$ joins the centre of $b$ to the centre of the sole triangle
it borders.
%%%%%%%%%%%%%%%%%%%%%%%%% 
\begin{figure}[h]
  \centering
  \begin{tikzpicture}[scale=1.0,line width=1.5pt]
    \coordinate (s1) at (0,0);
    \coordinate (s2) at (1,0);
    \coordinate (s3) at (0.3,0.8);
    \coordinate (s4) at (1.4,1.3);
    \coordinate (s5) at (0.6,1.5);
    \coordinate (s6) at (-0.5,1.2);
    \coordinate (s7) at (2.0,0.3);
    \coordinate (s8) at (2.2,1.3);
    %%%
    \coordinate (c1) at (0.4,0.27);  % 1,2,3
    \coordinate (c2) at (1.0,0.7); % 2,4,3
    \coordinate (c3) at (0.8,1.2); % 4,5,3
    \coordinate (c4) at (0.13,1.2); % 6,5,3
    \coordinate (c5) at (-0.1,0.7); % 1,6,3
    \coordinate (c6) at (1.5,0.5); % 2,4,7
    \coordinate (c7) at (1.9,1.0); % 2,6,7
%%%%    
    \coordinate (e1) at (-0.25,0.6);
    \coordinate (e2) at (0.05,1.35); 
    \coordinate (e3) at (1.0,1.4); %
    \coordinate (e4) at (1.7,0.8); %
    \coordinate (e5) at (1.5,0.15); %
    \coordinate (e6) at (0.5,0); %
    \coordinate (e7) at (1.8,1.3); %
    \coordinate (e8) at (2.1,0.8); %
\draw[color=black] (s1) -- (s2) -- (s3) -- (s1);    
\draw[color=black] (s2) -- (s4) -- (s3);    
\draw[color=black] (s4) -- (s5) -- (s3);    
\draw[color=black] (s5) -- (s6) -- (s3);    
\draw[color=black] (s6) -- (s1);    
\draw[color=black] (s2) -- (s7) -- (s4);    
\draw[color=black] (s7) -- (s8) -- (s4);    
\draw[color=red] (c1) -- (c2);    
\draw[color=red] (c2) -- (c3);    
\draw[color=red] (c3) -- (c4);    
\draw[color=red] (c4) -- (c5);    
\draw[color=red] (c1) -- (c5);    
\draw[color=red] (c2) -- (c6);    
\draw[color=red] (c6) -- (c7);
%%%
\draw[color=red] (c5) -- (e1);    
\draw[color=red] (c4) -- (e2);    
\draw[color=red] (c3) -- (e3);    
%\draw[color=red] (c6) -- (e4);    
\draw[color=red] (c6) -- (e5);    
\draw[color=red] (c1) -- (e6);    
\draw[color=red] (c7) -- (e7);    
\draw[color=red] (c7) -- (e8);    
  \end{tikzpicture}
\caption{
A triangulated system. Bonds are black, dual edges red. 
}\label{fig:primal-dual}
\end{figure}
%%%%%%%%%%%%%%%%%%%%%%%%%
Fig.~\ref{fig:cylinders} depicts the second type extended slightly, for clarity.
Each bond $b$ intersects exactly one dual edge, namely $\dual{b}$.
For a nonplanar triangulated surface, consider both bonds and
dual edges as curved in such a way that this intersection rule holds.

A frustrated triangle must have an even number of satisfied bonds,
in any configuration, see (\ref{eq:FM=Sat}) below.
Let \Sat\ denote the set of satisfied bonds in some unspecified configuration;
up to global spin flip, that configuration can be recovered from \Sat.
Then, the set of dual edges $\dual{\Sat}$
is a collection of \hbox{end-to-end} connected paths of dual edges with no terminus
inside the system.
Such a path is a \textit{string}, and a collection thereof, a \textit{string diagram}.
Fig.~\ref{fig:cylinders}(a) illustrates.
This is a local condition.
% Is every string diagram $\dual{\Sat}$ for some configuration, or
Are there global constraints?
On a cylinder, some FM/AFM bonding patterns force $\dual{\Sat}$ to intersect
every circumference an even number of times.
Others require odd intersection number.
The statistical ensemble of spin configurations is in \hbox{two-to-one}
correspondence with string diagrams (for $\dual{\Sat}$) which satisfy this
global constraint.
Each additional hole on the surface (in a homology sense, section~\ref{sec:homology})
brings an additional constraint, which collectively define a mod 2 cohomology
class (section~\ref{sec:cohomology}).
Every cohomology class is realized by some FM/AFM bonding pattern.

\subsection{Loops}\label{sec:loops}
%\subsection{Strings}
\label{sec:strings}

Let $\Cyc{C}$ be a path of end-to-end-connected bonds.
Supplemented with the spins it connects, it looks like this:
$\sigma_0b_1\sigma_1b_2\cdots \sigma_{N-1}b_N\sigma_N$.
Now, if $\sigma_1 \neq \sigma_0$, then we say there is a \textit{flip}
across bond $b_1$. A ferromagnetic (FM) bond is satisfied if there is
no flip, and an antiferromagnetic (AFM) bond, if there is a flip, so
\begin{equation}
  \label{eq:C-flip}
  \Cyc{C}^{\mathrm{flip}}
  \eqpar \Cyc{C}_{\mathrm{AFM}}^{\Sat}
  + \Cyc{C}_{\FM}^{\mathsf{Unsat}},
\end{equation}
where $\Cyc{C}_{\FM}^{\mathsf{Unsat}}$
denotes the set of unsatisfied FM bonds in $\Cyc{C}$, and so forth.
For $\Cyc{A}$, $\Cyc{B}$ sets of bonds
(or, later, dual edges), $\Cyc{A}+\Cyc{B}$ is their ``sum mod 2'',
so a bond is in that sum if it is in one and only one of $\Cyc{A}$
and $\Cyc{B}$. The subscript on the equals sign is a reminder that
we are working mod 2.
In the particular case of (\ref{eq:C-flip}), this is irrelevant
since the sets being added are disjoint, but it establishes the habit.
Now, by adding $2\Cyc{C}_{\mathrm{AFM}}^{\mathsf{Unsat}}$ or
$2\Cyc{C}_{\FM}^{\Sat}$, both of which are 0 (mod 2), we obtain
\begin{eqnarray}
  \label{eq:C-flip-2}
  \Cyc{C}^{\mathrm{flip}}
&  \eqpar \Cyc{C}_{\mathrm{AFM}}  + \Cyc{C}^{\mathsf{Unsat}}
  \nonumber \\
&  \eqpar \Cyc{C}_{\FM}  + \Cyc{C}^{\Sat}.
\end{eqnarray}
If all spins and all bonds in the path are distinct,
we can find a configuration on it which satisfies any desired pattern of
un/satisfied bonds:
Simply run along the path and cumulatively choosing successive spins to
un/satisfy the previous bond as needed.
Suppose we satisfy all the bonds, and
detach $b_N$ from $\sigma_N$ and connect it to $\sigma_0$ instead.
The result is a loop with at most one unsatisfied bond.
In any event, if we go around a loop with a spin configuration, counting
flips, there must be an even number. That is, for a loop $\Cyc{L}$,
and denoting the cardinality of a set by vertical bars,
$|\Cyc{L}^{\mathrm{flip}}|\eqpar 0$.
Therefore, specializing (\ref{eq:C-flip-2}) for a loop, we have the
important identity
\begin{equation}
  \label{eq:FM=Sat}
|\Cyc{L}_{\FM}|  \eqpar |\Cyc{L}^{\Sat}|.
\end{equation}
This is equivalent to 
$|\Cyc{L}^{\mathsf{Unsat}}| \eqpar |\Cyc{L}_{\mathrm{AFM}}|$.

  \subsection{Mod 2 homology}\label{sec:homology}

The collections of bonds which will interest from now on have no
free ends, so they are loops or sums of loops.
The latter are also known as \textit{cycles}.
For two cycles $\Cyc{A}$ and $\Cyc{B}$,
\begin{equation}
  \label{eq:FM-homo}
|(\Cyc{A} +  \Cyc{B})_{\FM}| \eqpar |\Cyc{A}_{\FM}| + |\Cyc{B}_{\FM}|,
\end{equation}
since, mod 2, both sides just count the number of FM bonds in the two cycles taken
together. In algebraic terminology, this says that the map
$\Cyc{A} \mapsto |\Cyc{A}_{\FM}|$ from the set of cycles to $\Int_2$
is a homomorphism of abelian groups.

Now, if $\Cyc{L}$ is the geometric boundary of a collection of
elementary triangles, it follows from (\ref{eq:FM-homo}) that
$|\Cyc{L}_{\FM}| \eqpar 0$,
because every elementary triangle has an even number of FM bonds.
%%%%%%%%%%%%%%%%%%%%%%%%% 
\begin{figure}[h]
\centering
\begin{tikzpicture}
  \node (bdy) at (0,0) {\includegraphics[width=90mm]{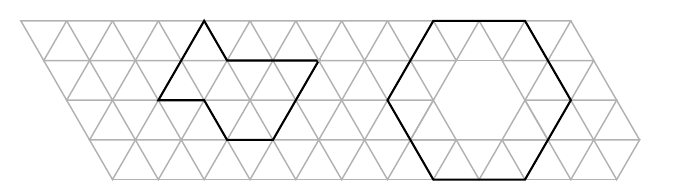}};
  \node at (-2.5,0.3) {\Large{$\mathscr{A}$}};
  \node at (0.6,0.4) {\Large{$\mathscr{B}$}};
  \node at (1.8,0.2) {\Large{$\mathscr{E}$}};
\end{tikzpicture}
\caption{Loop $\Cyc{A}$ is a boundary.
Neither loop $\Cyc{B}$ nor $\Cyc{E}$ is is a boundary, but their sum is so.
}\label{fig:bounding}
\end{figure}
%%%%%%%%%%%%%%%%%%%%%%%%%
In Fig.~\ref{fig:bounding}, $\Cyc{A}$ is a boundary ---
it is the sum of the triangles inside it. $\Cyc{B}$ is not a boundary,
however $\Cyc{B} + \Cyc{E}$ is one.
Whenever the difference --- or sum, mod 2 they are the same thing ---
of two cycles $\Cyc{A}$ and $\Cyc{B}$ is a boundary, they are called
\textit{homologous}, denoted $\Cyc{A} \sim \Cyc{B}$.
One also says that they belong to the same \textit{homology class},
denoted $[\Cyc{A}]$ (or $[\Cyc{B}]$).
Since $\Cyc{A}\sim\Cyc{A}'$ and $\Cyc{B}\sim\Cyc{B}'$ imply that
$\Cyc{A} + \Cyc{B} \sim \Cyc{A}' + \Cyc{B}'$, our mod 2 cycle addition
transfers to homology classes: $[\Cyc{A}] + [\Cyc{A}'] = [\Cyc{A}+\Cyc{A}']$.
With this addition, the homology classes are the elements of the (one-dimensional)
homology group of the system, $H_1$.
(For a general introduction to homology, see, e.g., Refs.
\cite{Frankel-GeoPhys,Giblin,Nakahara}.)

Homology is a geometrical relation, having nothing to do with
the FM or AFM nature of bonds. However, because elementary triangles
have an even number of FM bonds in a fully-frustrated system,
so does a sum of such triangles, so that
\begin{equation}
\Cyc{A} \sim \Cyc{B} \;\Rightarrow\;|\Cyc{A}_{\FM}| \eqpar  |\Cyc{B}_{\FM}|.
\end{equation}
Therefore, $[\Cyc{A}] \mapsto |\Cyc{A}_{\FM}|$ is well-defined, and since it
respects addition, as we have already seen, it is a group homomorphism
$H_1 \rightarrow \Int_2$.
The set of such homomorphisms is the (one-dimensional)
\textit{cohomology group} of the system, $H^1$. This is an abstract definition.
Fortunately, it has a simple geometrical representation
in terms of \textit{string diagrams}.

\subsection{String diagrams and mod 2 cohomology}\label{sec:cohomology}

Now consider $\dual{\FM}$, consisting of the duals of all the FM bonds.
Since each elementary triangle contains zero or two FM bonds,
$\dual{\FM}$ is a string diagram.
For any string diagram $\Str{A}$
and cycle $\Cyc{C}$, let $\Str{A} \bullet \Cyc{C}$ denote the number of their
intersections, mod 2.
(Note: script font for cycles, calligraphic for string diagrams.)
The map $\Cyc{C} \mapsto \Str{A} \bullet \Cyc{C}$ is an element of $H^1$,
that is, it is an additive map (homomorphism) which sends boundaries to zero.
The first property holds because counting intersections
is additive, and the second because the string diagram $\Str{A}$ has no
ends inside the system, so $\Str{A} \bullet \Cyc{T} \eqpar 0$ for each
loop $\Cyc{T}$ around an elementary triangle.
We call two string diagrams $\Str{A}$ and $\Str{B}$ \textit{cohomologous},
written $\Str{A}\sim\Str{B}$,
just in case $\Str{A}\bullet\Cyc{C} \eqpar \Str{B}\bullet\Cyc{C}$ for every
cycle $\Cyc{C}$.
(It suffices to test one loop in each homology class.)

% This is their \textit{intersection number}.
Returning to FM, 
since the FM bonds of $\Cyc{C}$ are exactly those intersected by
$\dual{\FM}$, it follows that
\begin{equation}
|\Cyc{C}_{\FM}| \eqpar \dual{\FM}\bullet \Cyc{C}.
\end{equation}
In exactly the same way, we obtain
$|\Cyc{C}^{\Sat}| \eqpar \dual{\Sat}\bullet \Cyc{C}$,
for any configuration. Finally, since 
$|\Cyc{C}^{\Sat}| \eqpar |\Cyc{C}_{\FM}|$, it follows that
$\dual{\Sat} \sim \dual{\FM}$.
So, in order that a string diagram $\Str{D}$ be $\dual{\Sat}$ for some
configuration, it is \textit{necessary} that $\Str{D}$ be cohomologous to
$\dual{\FM}$. This is a nonlocal condition.
To replace sum over configurations with sum over $\dual{\Sat}$
string diagrams, it is crucial to know if this condition is \textit{necessary}.
Section \ref{sec:all-in-class} shows that it is.
A related question is whether every cohomology class is realized by some
$\dual{\FM}$. Section \ref{sec:all-cohom} shows that this is also true.

\subsection{Every string diagram cohomologous to $\dual{\FM}$
  is $\dual{\Sat}$ for some configuration.}\label{sec:all-in-class}

Let $\dual{D}$ be a string diagram cohomologous to $\dual{\FM}$.
We show how to construct a spin configuration such that ${D} = \Sat$.
First, obtain a spanning tree $T$. This is a collection of bonds containing no
loops (tree) and such that every spin is touched (spanning).
[To construct one, start with any tree. If there is a site not touched by it,
connect it by a simple path of bonds not already in the tree. Continue until all
spins are touched. The procedure terminates since the system is finite.]
By a simple generalization of the procedure in Sec.~\ref{sec:loops},
find a configuration $\sigma$ such that a bond in $T$ is satisfied precisely
when it is in $D$. 
Let $b$ be a bond not in $T$. Necessarily, the spins it joins are touched by $T$,
so there is a loop $\Cyc{L}$ made of $b$ and some bonds in $T$.
Now, 
\hbox{$\dual{D}\bullet \Cyc{L} \eqpar \dual{\FM}\bullet \Cyc{L} \eqpar \dual{\Sat}\bullet \Cyc{L}$}, the first equality by hypothesis and the second by (\ref{eq:FM=Sat}).
But $D$ and \Sat agree everywhere on $\Cyc{L}$ except possibly $b$, they must
agree there as well, otherwise the equality could not hold.
Thus, $D = \Sat$ for the constructed configuration.

\subsection{Every cohomology class is represented by a string diagram}\label{sec:all-cohom}

%%%%%%%%%%%%%%%%%%%%%%%%% 
\begin{figure}[h]
  \centering
\includegraphics[width=70mm]{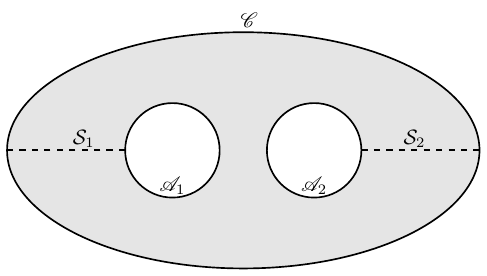}  
\caption{
  The edges $\Cyc{A}_1$ and $\Cyc{A}_2$ of the holes generate $H_1$.
$\dual{\FM}\bullet\Str{A}_i$ can be
independently toggled between $0$ and $1$ by adding $\Str{S}_i$ to $\dual{\FM}$.
}\label{fig:bond-cohomology}
\end{figure}
%%%%%%%%%%%%%%%%%%%%%%%%%

That every cohomology class is realizable as $\dual{\FM}$ can be proved 
by linear algebraic reasoning and finiteness of the set of bonds, since $\Int_2$ is a field.
Since we will not need this result in the following, we omit a general proof, and
consider the special case of surfaces which can be homeomorphically embedded in the plane.
Fig.~\ref{fig:bond-cohomology} shows one such. Topologically, a cylinder is a disk with
one hole, so it is covered as well.
The main point is very simple. Every cycle is homologous to a sum of the boundaries of
zero or more holes. We can toggle the value of each $\dual{\FM}\bullet\Cyc{A}_i$
independently by adding the string $\Str{S}_i$ to $\dual{\FM}$.
%%%%%%%%%%%%%%

%%%%%%%%%%%%%%%
\section{Ground configurations}\label{sec:ground}

Now we leave behind general triangulated surfaces to focus on regular triangular
lattices wrapped on cylinders. To the extent that we are interested in the
thermodynamic limit, global topology ought not to matter, anyway.
We modify the string representation of configurations developed in the previous
section into one especially suitable at low temperature, and ripe for
reinterpretation in terms of (almost completely) noninteracting fermions. 

To help distinguish the string diagrams already introduced from the modified
ones developed in this section, we refer to the former as \textit{absolute}
string diagrams (and the latter, naturally, as \textit{relative} string diagrams).
After this section, we work only with the relative ones, and will then
drop the adjective.
This terminology is potentially misleading insofar as it is really the
interpretation which makes the diagram \textit{absolute} or \textit{relative}.
\textit{String diagram} in the abstract is still perfectly well-defined.

\subsection{Local condition for ground configurations}

Since every link of a configurational absolute string diagram $\Cal{D}$
represents a satisfied bond, those which correspond to ground configurations are
those with maximal length $|\Cal{D}|$.
An unpleasant feature of this characterization is its nonlocal character;
we have to count up the total length. Is there a local condition?
%%%%%%%%%%%%%%%%%%%%%%%%% 
\begin{figure}[h]
  \centering
\includegraphics[width=35mm]{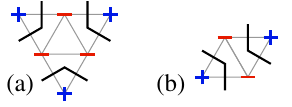}  
\caption{(a) A ground configuration containing a triangle with all three of its bonds
  frustrated. This is a down-triangle-type system.
  (b) A system which is neither up-triangle-type nor down-triangle-type.
Millane and coworkers~\cite{Millane+Blakeley-04,Millane+Clare-06,Blakeley+Millane-06}
have investigated such irregular systems in depth.
}\label{fig:tiny}
\end{figure}
%%%%%%%%%%%%%%%%%%%%%%%%%

A sort of folklore says that in a TAFIM system ground state
configuration, \textit{every} elementary triangle has exactly two satisfied bonds.
That is a local condition. It corresponds to absolute diagrams with string going through
every triangle.
Unfortunately, in general, it is not quite right~\cite{Blakeley+Millane-06,Millane+Clare-06}.
The reason is that in a finite system, interior bonds are shared by
an up-pointing triangle (\uptri, \textit{up-triangle} for short) and a
down-pointing triangle
(\dntri, \textit{down-triangle}), but those on the system boundary are not.
This is illustrated on Fig.~\ref{fig:tiny}, which shows both spins
and the corresponding absolute string diagram assuming all bonds AFM.
In the little system of Fig.~\ref{fig:tiny}(a), every bond belongs to
a down-triangle. Hence the lowest energy possible has a string through every
down-triangle. All the bonds are then accounted for, so what happens in the
up-triangle is irrelevant. This generalizes to any system for which every
bond belongs to a down-triangle; we call them down-triangle-type.
Similarly, there are up-triangle-type-type systems; they are just down-triangle-type
systems upside-down. Toroidal systems belong to both types,
while others, such as that in Fig.~\ref{fig:tiny}(b), do not fall
into either category. This last category are the hardest to analyze.
In this paper, we restrict ourselves
to down-triangle-type systems. For cylinders, that means we draw them with
a smooth edge on top and a serrated edge on the bottom.

The point of this restriction is that we have a local criterion for ground
configurations of a down-triangle-type system is easy.
The motifs \usebox{\downleft} and \usebox{\downright}
are the lowest energy configurations for a down-triangle.
The motif \usebox{\emptydn} costs energy $4J$ relative to these.
(Complete with coupling strength, bond energy is $\pm J\sigma\sigma'$.)
Hence, measuring energy relative to the ground state, the total energy of
a down-triangle-type system is $4J$ times the number of \usebox{\emptydn}'s.
In particular, ground configurations have no empty \textit{down-triangles}.

%%%%%%%%%%%% 
\subsection{Relative string diagrams}\label{sec:relative-diagrams}

Now we do some mod 2 arithmetic with string diagrams.
For two string diagrams $\Str{D}$ and $\Str{D}'$,
$\Str{D} \eqpar (\Str{D} + \Str{D}') + \Str{D}'$.
Therefore, if we had a \textit{reference} diagram $\Str{D}_{\mathrm{ref}}$,
the transformation \hbox{$\Str{D} \mapsto \Str{D} + \Str{D}_{\mathrm{ref}}$} is
a one-to-one mapping of the set of string diagrams onto itself.
Perhaps such a transformation can give a more useful configurational
representation.
Conditions can be given which leave only one nontrivial candidate for
$\Cal{D}_{\mathrm{ref}}$.
For instance, we would like it to work uniformly for all cylinders.
This really means that there is some sort of template from which we get
$\Cal{D}_{\mathrm{ref}}$ for every cylinder, and therefore conflate them by
saying ``the'' reference diagram. Also, it should respect the translation
invariance of the lattice, otherwise it would be very complicated to work with
(and would have difficulty satisfying the first condition).
It is probably also helpful if $\Cal{D}_{\mathrm{ref}}\bullet \Cyc{L}\eqpar 0$
for every loop $\Cyc{L}$. Then the intersection number of valid absolute and
relative diagrams with a circumferential loop will be the same.
These considerations identify
%  a loop $\Cyc{L}$, 
%  \hbox{$(\Str{D} + \Str{D}')\bullet\Cyc{L}
%    \eqpar \Str{D}\bullet\Cyc{L} + \Str{D}'\bullet\Cyc{L}$}
%  
%  \begin{equation}
%    \Cal{L} \cap (\Str{D} + \Str{D}_{\mathrm{ref}})
%    \eqpar  
%    \Cal{L} \cap \Str{D} + \Cal{L} \cap \Str{D}_{\mathrm{ref}}.
%  \end{equation}
%  the ``$+\Cal{D}_{\mathrm{ref}}$'' transformation preserves the class of regular diagrams
%  if
%  \begin{equation}
%    \label{eq:preserves-regularity}
%    \Cal{L} \cap \Cal{D}_{\mathrm{ref}} \eqpar 0,
%    \;\mathrm{for}\;\mathrm{all}\;\mathrm{loops}\; \mathcal{L}.
%  \end{equation}
%  If, in addition, we want a reference diagram which has the full periodicity
%  of the underlying lattice, and works for all cylinders, there is really only
%  one choice.
%  
the diagram $\Cal{D}_{\mathrm{zigzag}}$ in Fig.~\ref{fig:cylinders}(b),
containing only the motifs \usebox{\annihilation} and \usebox{\creation},
and zig-zagging horizontally across the plane, or around the cylinder circumference.
However, being singled out like this is not particularly important.
The real question is whether the transformation is useful, and that remains to be shown.

%%%%%%%%%%%%%%5
\begin{figure}[h]
  \centering
  \includegraphics[width=60mm]{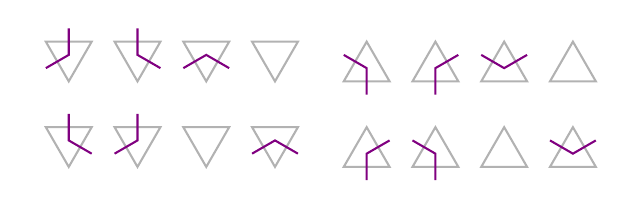}
  \caption{Motif transformations effected by $+\Cal{D}_{\mathrm{zigzag}}$
(Fig.~\ref{fig:cylinders}(b)). Since the operation is involutive,
    one can read from top to bottom or bottom to top.}
  \label{fig:+Dref}
\end{figure}
%%%%%%%%%%%%%%5
The action of the operation $+\Cal{D}_{\mathrm{zigzag}}$ is purely local
and illustrated in Fig.~\ref{fig:+Dref}. The important thing is that
the motif with an energy cost changes from \usebox{\emptydn} to \usebox{\creation}.
Thus, since that motif is necessary to have string loops in the diagram,
relative diagrams for ground configurations do not contain loops.
Supplemented with boundary conditions at the cylinder top and bottom,
this guarantees that every horizontal slice through the cylinder sees
the same number of strings.
The number of strings entering the top is straighforwardly the number of
satisfied bonds along that edge. At the bottom, a little trick is helpful.
We imagine ``phantom'' bonds along the lower edge, which do not count toward
the energy, but allow to count strings exiting.

%%%%%%%%%%%%%%%%%%%%%%%%%% 
\section{From strings to fermions}\label{sec:fermions} 

Henceforth, \textit{string diagram} means
\textit{configurational string diagrams relative to $\Cal{D}_{\mathrm{zigzag}}$},
unless otherwise indicated.

\subsection{Strings as particle worldlines.}

%%%%%%%%%%%%%%%%%%%%%%%%%%%%%
\begin{figure}[h]
  \centering
  \includegraphics[width=75mm]{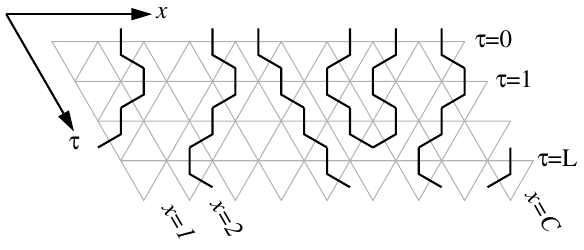}
  \caption{Strings as worldlines.
  }
  \label{fig:space-and-time}
\end{figure}
%%%%%%%%%%%%%%%%%%%%%%%%%%%%%

We can think of string diagrams as
depictions of particle worldlines, as suggested by the labelling
in Fig.~\ref{fig:space-and-time}.
  Here we make a choice to keep $x$ constant on rows going to the
  lower right rather than the lower left. This will lead to an asymmetry
  in the hopping later. It is necessary to make one choice or the other
  in order that allowed positions be independent of (integer) time.
Circumferential rings of bonds correspond to successive integer
times $\tau=0,1,\ldots$, and the bond centres correspond to integer
positions modulo $C$;
$x\in\Int_C = \{1,2,\ldots,C\}$, so $0\equiv C$ and $C+1\equiv 1$.
In this particle interpretation, the motif \usebox{\annihilation}
corresponds to annihilation of pairs of neighboring particles.
For general boundary conditions, this may occur in
zero-temperature diagrams, but \usebox{\creation}, representing pair
creation, may not. This asymmetry is because our systems are down-type,
and by convention the flat end corresponds to ``earlier''.
We also recall that the reason we insist on a definite type is that
it gives a local criterion for a string diagram representing a ground
configuration. Without that, it is not clear how or whether we could
obtain a local Hamiltonian description.

Unless otherwise noted, we assume boundary conditions are such as to
force the same number of particles exiting at the bottom as entering at the top,
such as simple periodic boundary conditions.
In that case, annihilation events are strictly forbidden at zero temperature.

\subsection{Fermions}\label{sec:Fock-space}

If one of our $\NN$
particles is at position $x$ at time $\tau$, then in the next time step,
it can remain where it is, or move to $x-1$. However, distinct particles
cannot occupy the same position.
Because our particles are moving in one dimension, this hard-core
interaction can be implemented by declaring the particles to be fermions.
Thus, we consider
operators $c_n^\dag$, $n=1,\ldots,C$, obeying the anticommutation
relations $\{c_n,c_m\} = 0$, $\{c_n^\dag,c_m\}=\delta_{n,m}$ and
the Fock space generated by their action 
on the vacuum state $\ket{\Vac}$ (containing no particles).

A particle configuration $X=(x_1,\ldots,x_\NN)$ is a set of positions,
by convention in ascending order: $1 \le x_1 < x_2 < \ldots < x_\NN \le C$.
This configuration is identified with the Fock state vector
\begin{equation}
\ket{X} \defeq c_{x_1}^\dagger \cdots c_{x_\NN}^\dagger \ket{\Vac}.
\label{eq:configuration-to-Fock-space}
\end{equation}
In the original spin model, probabilistic mixtures of different configurations
makes sense, and we can consider the classical state space ${\mathcal S}$ to
consist of assignments $X \mapsto p(X)$ of probabilities to configurations.
Such a state is associated with a Fock state vector, $\sum p(X)\ket{X}$,
by linear extension.
Ignoring normalization, the classical state space
$\mathcal S$ consists effectively of linear combinations of the basis vectors
$\ket{X}$ with non-negative coefficients,
  whereas states in the Fock space
  may involve this basis with complex coefficients.
This distinction is important for the interpretation of the results.
The transfer matrix (equivalently, the Hamiltonian) of the fermion system has a full
set of eigenstates in Fock space. If such an eigenstate $\ket{i}$ made
sense as a \textit{spin} boundary condition, that is belonged to $\mathcal S$,
then that state could be forced through the entire length of the cylinder.
In fact, only a few of the eigenstates belong to $\mathcal S$.

\subsection{Particle-number conserving transfer matrix.}\label{sec:xfer_0}

\subsubsection{real space}

Now we construct the particle-number conserving ring-to-ring transfer matrix $\xfer_0$.
Modifications for nonzero temperature will be taken up in Section~\ref{sec:T>0}.

A particle at $x$ at time $\tau$ has two options for its position at time $\tau+1$:
$x$ or $x-1$ (see Fig.~\ref{fig:cylinders}c).
Thus, $\xfer_0$ acts as
\begin{equation}
\xfer_0\, c_{x_1}^\dagger \cdots c_{x_\NN}^\dagger \ket{\Vac}
= (c_{x_1-1}^\dagger + c_{x_1}^\dagger) \cdots 
(c_{x_\NN-1}^\dagger + c_{x_\NN}^\dagger) 
\ket{\Vac}.
\nonumber
\end{equation}
The right-hand side contains every possibility, counted once.
Because the fermions cannot occupy the same location, their order is maintained,
up to cycling around the ring.
$\xfer_0$ is thus determined by the conditions
\begin{equation}
\xfer_0 \ket{\Vac} = \ket{\Vac},
\qquad
\xfer_0 c_x^\dagger = (c_{x-1}^\dagger + c_{x}^\dagger) \xfer_0.
\label{eq:T0-conditions-x}
\end{equation}
There is a subtle point, however, connected with the possibility of moving from
$x=1$ to $x = 0 \equiv C$ or vice versa.
Since
\begin{equation}
c_{x_1}^\dagger \cdots c_{x_\NN}^\dagger \ket{\Vac} 
= (-1)^{\NN-1}c_{x_2}^\dagger \cdots c_{x_\NN}^\dagger c_{x_{1}}^\dagger \ket{\Vac},
\end{equation}
we must take
%%% PEL
\begin{equation}
   \label{eq:N-parity-and-c-BC}
   c_{j+C} =
   \left\{
     \begin{array}{lr}
c_{j}  & \NN \, \mathrm{odd}, \\
-c_{j} & \NN\, \mathrm{even}.
     \end{array}
     \right.
\end{equation}
Since only the parity of $\NN$ matters here, there will be no problem with this
rule even with pair creation/annihilation.
Conditions (\ref{eq:T0-conditions-x}) need to solved with that constraint imposed.
However, it makes no sense to do that directly in position space. 

\subsubsection{$k$-space}\label{sec:k-space}

We pass to $k$-space, using
\begin{equation}
c(k)^\dagger = \frac{1}{\sqrt{C}}\sum_{n\in\Latt} e^{ikn} c_n^\dagger,  
\end{equation}
where $k$ is in the Brillouin zone
%%% PEL
%   \begin{equation}
%   \label{eq:BZ}
%   \BZ = 
%   \begin{cases}
%   \frac{2\Int}{C}\pi \cap (-\pi,\pi], & \NN \, \mathrm{odd}
%   \\
%   \frac{2\Int+1}{C}\pi \cap (-\pi,\pi], & \NN\, \mathrm{even},
%   \end{cases}
%   \end{equation}
\begin{equation}
\label{eq:BZ}
\BZ = 
\left\{
  \def\arraystretch{1.5}
  \begin{array}{ll}
\frac{2\Int}{C}\pi \cap (-\pi,\pi], & \NN \, \mathrm{odd}
\\
\frac{2\Int+1}{C}\pi \cap (-\pi,\pi], & \NN\, \mathrm{even}.
  \end{array}
\right.
\end{equation}
The allowed values of $k$ here are dictated by (\ref{eq:N-parity-and-c-BC}).

Condition (\ref{eq:T0-conditions-x}) on $\xfer_0$ is equivalent to
\begin{equation}
\xfer_0 c(q)^\dagger 
= {C}^{-1/2}\sum_{x \in\Latt} e^{iq x} \xfer_0 c_x^\dagger 
= \left(2\cos \frac{q}{2}\right) e^{iq/2} {c}(q)^\dagger \xfer_0.
  \label{eq:T-c-dag}
\end{equation}
From this, we deduce that
\begin{equation}
  \label{eq:xfer-k-factors-1}
\xfer_0 = \prod_{k \in {\BZ}} 
\left[ 1 - n(k) + \left( 2\cos\frac{k}{2}\right) e^{ik/2}n(k)\right],
\end{equation}
in terms of the occupation number operators
\begin{equation}
  \label{eq:occupation-number}
n(k) \defeq c(k)^\dagger c(k).
\end{equation}
%%%%%%%%%%%%%%%%%%%%%
Finally, using
$(1-n(k)) + e^\alpha n(k) = e^{\alpha n(k)}$, rewrite (\ref{eq:xfer-k-factors-1}) as
\begin{equation}
\label{eq:T-exponential-form}
\xfer_0
= e^{iP/2} e^{-{H}_0},  
\end{equation}
where
\begin{equation}
\label{eq:HP}
{H}_0 = \sum_{q\in \BZ} {\varepsilon}(q) n(q)
          \quad \mathrm{and} \quad
P = \sum_{q\in\BZ} q\, n(q),
\end{equation}
with
\begin{equation}
{\varepsilon}(q) = -\ln\left(2\cos\frac{q}{2}\right).
\label{eq:mode-energy}
\end{equation}
As the notation suggests, $H_0$ and $P$ can be considered a Hamiltonian and total
momentum operator, respectively for the fermion system.
Ignoring the $e^{iP/2}$ factor, occurence of which
will be elucidated in section~\ref{sec:free-energies-and-entropies},
(\ref{eq:T-exponential-form}) says that the transfer matrix
$\xfer_0$ evolves the system for an imaginary time \hbox{$\Delta t = -i$}.
In fact, the $e^{iP/2}$ factor is not particularly important in the thermodynamic
limit $L\to\infty$.
%%%%%%%%%%%%%%%%%%%
\section{Equilibrium TAFIM ground states and their entropy densities}\label{sec:entropy}

An equilibrium macrostate of a finite lattice system is determined by the temperature,
the Hamiltonian, and boundary conditions.
For systems, such as TAFIM, with many ground configurations, 
this concept is nontrivial also at zero temperature. The equilibrium macrostate is
a uniformly weighted statistical mixture of all ground configurations compatible
with the boundary conditions.
The standard way to define an equilibrium macrostate for an infinite system
is to consider a sequence of increasing finite systems equipped with boundary
conditions such that expectations for all observables in all finite regions
(``in the bulk'') converge. The limiting values define the equilibrium macrostate.
We can apply this to ground states as well.
An alternative view, due to Dobrushin, Lanford, and Ruelle, works directly
with the infinite system.
A probability distribution $\mathsf{Pr}$ on configurations is an equilbrium
state if it satisfies this consistency condition:
For any finite subsystem $\Lambda$, the distribution of spins in
$\Lambda$ \textit{conditional} on the exterior is simply a Gibbs distribution.

Above the critical temperature,
two-dimensional ferromagnetic Ising models with zero magnetic field have a single
planar limit equilibrium macrostate (henceforth possibly shortened to simply ``equilbrium state'').
Below the critical temperature, including at zero temperature, there are precisely two.
The TAFIM, in contrast, has a unique equilibrium macrostate for all nonzero temperatures.
Asymptotically, boundary conditions do not matter. What about at zero temperature? 
Surpisingly, there is a \textit{two-dimensional continuum} of ground states,
characterized by the densities of frustrated bonds in each of the three orientations.
Those fractions must average one-third (equivalently, sum to one).
However, any combination satisfying that restriction can be selected by
boundary conditions even in the thermodynamic limit. 

The nature of the continuum of equilibrium macrostates has an interesting
description in the fermionic language.
As we have seen, fermion number, or \textit{charge} to use a more evocative
word is locally conserved. Thus, over long times,
every point of a ring system must experience the same the time-average current,
but the latter simply corresponds to
$\jj = \bar{\nn}(\usebox{\urline})  -  \bar{\nn}(\usebox{\ulline})$,
the local difference between density of bonds with the two non-circumferential
orientations.
This suggests that we will be dealing with eigenstates of $H_0$ with nonzero
momentum $P$, but we have already cautioned that such states cannot be selected
by spin boundary conditions. The resolution of this little puzzle is that
we must modify the Hamiltonian by what can be interpreted as an imaginary
vector potential.

\begin{figure}
\centering
\includegraphics[width=80mm]{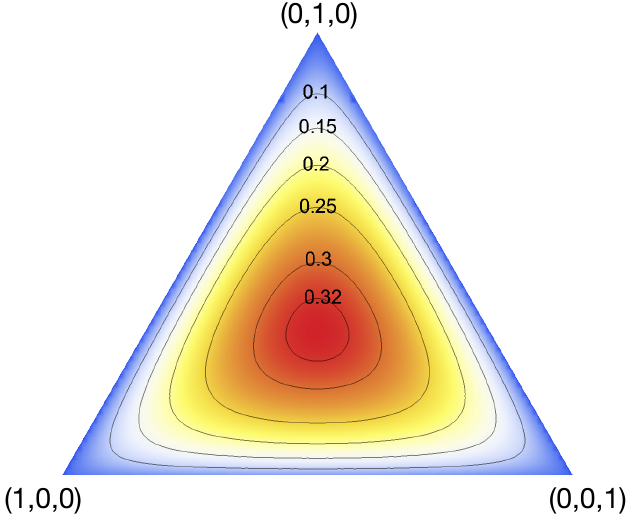}
\caption{
  Entropy density
  $s$
as function of unsatisfied bond fractions
$\bar{\nn}(\usebox{\horline})$,
$\bar{\nn}(\usebox{\urline})$,  
and $\bar{\nn}(\usebox{\ulline})$  
in the three orientations.
The vertices correspond to the extreme unsatisfied bond fractions --- all and
only bonds with one particular orientation unsatisfied.
Any point in the diagram is a convex sum of the vertices, so that
the point which is the vector sum
$\bar{\nn}(\usebox{\horline})(1,0,0) + \bar{\nn}(\usebox{\urline})(0,1,1)
+ \bar{\nn}(\usebox{\ulline})(0,0,1)$ of the vertices represents the state
with unsatisfied bond fractions
$\bar{\nn}(\usebox{\horline})$,
$\bar{\nn}(\usebox{\urline})$, and
$\bar{\nn}(\usebox{\ulline})$.
A slice through this plot along a line from a vertex to the midpoint of
the opposite edge reproduces the curve of Fig.~\ref{fig:entropy-curve}.
\label{fig:entropy-surface}}  
\end{figure}

The entropy densities of all these infinite-system equilibrium ground states vary,
as plotted in Fig.~\ref{fig:entropy-surface}.
There is a unique such with maximal entropy density, namely that with
one-third of the bonds of each orientation unsatisfied.
It is the limit of the unique $T>0$ equilibrium macrostates.
This does not imply that the others are completely irrelevant to
$T>0$ physics, however. We crudely estimate that at low but nonzero
temperature, something very close to one of the equilibrium ground
states can be boundary-condition stabilized in a finite system with
linear extent of order $e^{4 J/T}$.
Artificial spin ice systems\cite{ASI-Nature,Schiffer+Nisoli-21}
are a good place to look for this sort of control by boundary conditions.
The interactions in these systems of nanoscale magnetic particles can be
easily adjusted to a range where room temperature is, effectively, zero,
their configurations can be studied in complete detail, and individual
particles can be magnetically manipulated.

%%%%%%%%%%%%%%%%%%%%%%%%% 
%%%%%%%%%%%%%%%%%%%%%%%%%

\subsection{Cylinders}\label{sec:cylinder-entropy}

\subsubsection{Free energies and free entropies.}\label{sec:free-energies-and-entropies}

From the results of the previous section, 
we conclude that the \textit{twisted} partition function
\begin{eqnarray}
%  \tilde{Z}_{C}(\frac{\small{1}}{\tiny{L}},\NN)
\tilde{Z}_{C}({L}^{-1},\NN)
  &=\Tr_\NN \left( e^{iPL/2} e^{-H_0 L} \right)
    \label{eq:Zferm-N}
\end{eqnarray}
of $\NN$ fermions on a ring of circumference $C$ at temperature $T=L^{-1}$
is equal to the partition function $\partfn_{C,L}(0,\NN)$ of the
zero-temperature TAFIM on a $(C,L)$-torus, with boundary conditions forcing
$\NN$ satisfied bonds on every circumference.
A note about the notation: italic $Z$ denotes a partition function of the
fermion system, calligraphic $\mathcal Z$, one for a spin system.
A subtle point here, indicated by the adjective `twisted',
is the state rotation implemented by the operator $e^{iPL/2}$ inside the trace.
Refering to Fig.~\ref{fig:cylinders} or \ref{fig:space-and-time}, the
explanation is as follows:
Because of the slant of constant-$x$ lines, $(x,\tau=L)$ is directly below
$(x+L/2,0)$ in the natural planar or cylindrical representation of
those figures.
We should therefore restrict $L$ to be even for the sake of closing the
system into a torus.
The twisting in the trace defining $\tilde{Z}$, indicated by the tilde,
enforces equality of the state at $\tau=L$ rotated forward by $L/2$ with the state
at $\tau=0$. 
This is twisting with respect to the $x$ coordinates is precisely what is needed
to achieve the identification of the top and bottom of the cylinder
in the natural geometrically untwisted way.

  Now, we obtain the correspondence between thermodynamic potentials of
  the TAFIM system and its fermionic representation, namely
\begin{equation}
  \label{eq:fermion-f=spin-phi}
e^{-LC f_C\left({L}^{-1},\nn \right)}
 = \tilde{Z}_{C}({L}^{-1},\NN)
%    \nonumber \\
  = \partfn_{C,L}(0,\NN)
%    \nonumber \\
 = e^{\Nspin \phi_{C,L}(0,\nn)}.
\end{equation}
On the left, we have the partition function of the fermion system
at (nonzero) temperature $L^{-1}$, and particle density
\begin{equation}
\nn = C^{-1} {\NN}.
\end{equation}
$f_C\left({L}^{-1},\nn \right)$ is its free {energy} density.
The TAFIM is at zero temperature, so we cannot have a bare
factor of $T^{-1}$ in the partition function exponent.
Therefore we absorb it by using the Massieu potential,
generally defined as $\phi = s - \beta e$, with $e$ is energy density.
Since we have taken our energy zero to be the ground state energy,
we just have $\phi(0,\nn) = s$, the entropy per spin. 

The relation (\ref{eq:fermion-f=spin-phi}) between the spin system and
the fermion system continues to hold for nonzero temperature $T$.
In that case, $\NN$ is understood as referring to boundary conditions only,
since for $T > 0$ these do not control the density in the bulk.
Also, the equivalent fermion Hamiltonian becomes $T$ dependent.
That situation is investigated in Section \ref{sec:T>0}.

The fermion free energy is awkward to evaluate in the
canonical ensemble, except in the zero-temperature limit $L\to\infty$,
where it becomes simply the energy of a filled Fermi sea:
\begin{equation}
s(\nn)
= - f_C(0,\nn) =
  -{\textstyle\frac{1}{C}}\sum_{|k| < k_F} \varepsilon(k)
%\nonumber \\
 \,\stackrel{C\to\infty}{\longrightarrow}\,
  -\int_0^{k_F} \varepsilon(k)\, \frac{dk}{\pi},
\label{eq:s(n)-integral}
\end{equation}
with the Fermi momentum
\begin{equation}
  \label{eq:kF}
k_F = \pi \nn.
\end{equation}

%%%%%%%%%%%%%%%
\begin{figure}\centering
\includegraphics[width=80mm]{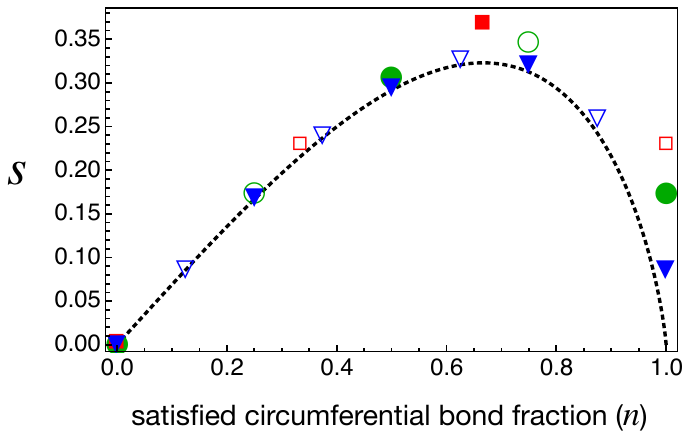}
\caption{
Zero-temperature entropy per spin of the cylindrical TAFIM as function
of satisfied circumferential bond fraction
(cf. Dhar \textit{et al.}\cite{Dhar+Chaudhuri+Dasgupta-00}).
Solid curve: $C\to\infty$ limit, given by the formula in Eq.~(\ref{eq:s(n)-with-Bernoulli}).
Symbols: exact numerical results for circumference 3 ($\square$), 4 ($\diamondsuit$) 
and 8 ($\triangledown$).
Solid (open) symbols denote periodic (antiperiodic) boundary conditions.
Except at $n=1$, for which entropy density is exactly $(\ln 2)/C$,
the $C=\infty$ limit is nearly attained already for $C=8$.
\label{fig:entropy-curve}}  
\end{figure}
%%%%%%%%%%%%%%%%%%%%%%
Fig.~\ref{fig:entropy-curve} shows $s(\nn)$, together with numerical results for $C=3,4,8$.
Evidently, for most densities, $C=8$ is already ``large''.
The maximum of $s(\nn)$ is attained at $n={2}/{3}$, %$\bar{\nn}(\usebox{\horline})=1/3$.
This is as expected, since under open boundary conditions, 
we would expect 1/3 of the bonds of \textit{each\/} orientation to be
unsatisfied, at least in the limit $C,L\to\infty$.

\subsubsection{Series expansion of $s(\nn)$.}\label{sec:entropy-series}

The equation (\ref{eq:s(n)-integral}) for the $C,L\to\infty$ limit of
the TAFIM entropy density is pleasant for its simple connection to a familiar problem.
Section~\ref{sec:integrals} derives a series expansion, the first few
terms of which are
\begin{equation}
\label{eq:entropy-series-2}
{\kern -6em}  s(\nn)
  =  -{\bar{\nn}}\ln \left({\frac{2}{e}}\sin{\frac{\pi\bar{\nn}}{2}} \right)
  -  \frac{\pi^2\bar{\nn}^3}{36}
     - \frac{\pi^4\bar{\nn}^5}{3600}
    - \frac{\pi^6\bar{\nn}^7}{211680}
    - \frac{\pi^8\bar{\nn}^9}{10886400}
- R_{11}(\bar{\nn}).
\end{equation}
Here,
\begin{equation}
\bar{\nn} \defeq 1-\nn,
\end{equation}
and the remainder satisfies $0 < R_{11}(\bar{\nn}) < 10^{-4}$.

The maximum of $s$ is of particular importance.
The only macrostates accessible by taking the thermodynamic limit first,
and then the limit $T\to 0$, are those with maximum entropy density\cite{Aizenman+Lieb-81}.
Assuming no spontaneous symmetry breaking,
this maximum must occur for $\nn=2/3$, because the densities of unsatisfied bonds
with each of the three orientations must be equal and sum to one.
From just the few terms in the formula (\ref{eq:entropy-series-2}) and the bound
$R_{11}(1/3) < 2\times 10^{-9}$, the ordinary residual entropy density is
 \begin{equation}
  \label{eq:residual-entropy}
\lim_{T\to 0} s
= s\paren{ {\textstyle\frac{2}{3}} }
= 0.3230659(9),
\end{equation}
extending the value reported by Wannier\cite{Wannier_erratum} by a couple of decimal places.

\subsubsection{Remarks.}\label{sec:BCs}

The constraint on the number of satisfied bonds arounds the circumference
is controllable by boundary conditions for a cylinder;
if boundary conditions enforce $\NN = \nn C$ satisfied bonds at top and
botton, then the same is true for every circumference.
However, just control of $\nn$ globally does not immediately imply that
it can be controlled in macroscopically large but finite regions for a
sequence of cylinders with $C,L\to\infty$.
We must rule out phase separation, which would result in
macroscopically large regions with values of $\nn$ smaller or larger
than the global value. That this does not happen is demonstrated by
the strict concavity of $s(\nn)$. Phase separation would give flat
parts of the graph.

The transfer operator (\ref{eq:T-exponential-form}) conserves other
quantities besides particle number, for example momentum.
This raises the question whether there is an even richer variety
of thermodynamic limit states. Sadly, the answer is no.
The fermion language is more expressive than the spin language,
as touched upon in section \ref{sec:Fock-space}.
Every spin boundary condition corresponds to a Fock space state
with a zero momentum component, which inevitably dominates as $L\to\infty$.

%%%%%%%%%%%%%%%%%%% 
\subsection{All planar limit equilibrium macrostates}\label{sec:planar-entropy}

This subsection determines \textit{all} the planar thermodynamic limit equilibrium
states and their entropy densities. Fig.~\ref{fig:entropy-curve} represents only
some of these.

\subsubsection{Boundary controllability of satisfied bond densities.}

The first step is to show that the global densities, 
$\nn(\usebox{\horline})$, $\nn(\usebox{\urline})$, and $\nn(\usebox{\ulline})$,
of \textit{satisfied} bonds in the three distinct orientations are controllable 
by boundary conditions, subject only to the constraint
(since overall, $2/3$ of the bonds are satisfied)
\begin{equation}
{\nn}(\usebox{\urline})  + {\nn}(\usebox{\ulline}) + {\nn}(\usebox{\horline}) = 2.
\end{equation}
To see this, refer to our standard representation of a finite lattice,
as in Fig.~\ref{fig:space-and-time}. Assume we have periodic boundary
conditions top-to-bottom, restricted to configurations with
$N(\usebox{\horline})$ satisfied bonds, and, as usual, periodic
boundary conditions left-to-right. Now, the fact that every horizontal
slice has $N(\usebox{\horline})$ satisfied bonds holds for each configuration,
so it is still true if we add a restriction that there are
$N(\usebox{\ulline})$ satisfied bonds along the left (and thus right) edge.
But now the situation is essentially symmetric; rotating the figure
interchanges top/bottom with left/right. Conclude that, for every configuration,
every row of bonds parallel to the left and right edges also contains
$N(\usebox{\ulline})$ satisfied bonds.
This shows that \textit{global} control is possible.
Strict concavity of the entropy density will show that in the thermodynamic
limit equilibrium macrostates, the satisfied bond densities in every macroscopic
region agree with the boundary condition induced global density.

\subsubsection{Bond weights and grand canonical ensemble.}

In the present situation, it is much easier to do the actual calculations
in what could be called a grand canonical ensembles.
Thus, we couple control fields $\mu$ and $\alpha$ to the numbers
$N(\usebox{\horline})$, $N(\usebox{\ulline})$, and $N(\usebox{\urline})$ 
of satisfied bonds with the indicated orientations, 
and consider the partition function
\begin{equation}
  \label{eq:Z-mu-alpha}
\partfn_{C,L}(0,\mu,\alpha)
\defeq
\sum_{\mathrm{ground}\;\mathrm{diagrams}} {\kern-1em}
e^{
  \mu N(\usebox{\horline}) + \alpha  [N(\usebox{\ulline}) - N(\usebox{\urline})]
}.
\end{equation}
on the $(C,L)$ torus, i.e., with simple periodic boundary conditions.
The reason we have only two multipliers ($\mu,\alpha$) instead of three
is that there is a constraint. Exactly $2/3$ of the bonds in a ground
microstate are satisfied, so
\begin{equation}
\bar{\nn}(\usebox{\urline})  + \bar{\nn}(\usebox{\ulline})  = {\nn}(\usebox{\horline}),
\end{equation}
where $\nn(*)$ is the \textit{fraction} of orientation-$*$ bonds which are
satisfied, and $\bar{\nn}(*) = 1-\nn(*)$ is the fraction of \textit{unsatisfied}
bonds with the same orientation. Then, defining 
\begin{equation}
  \label{eq:j}
  \jj \defeq
  \bar{\nn}(\usebox{\urline})  -  \bar{\nn}(\usebox{\ulline})  
=  {\nn}(\usebox{\ulline})  -  {\nn}(\usebox{\urline}),  
\end{equation}
the exponent in (\ref{eq:Z-mu-alpha}) is rewritten as
\begin{equation}
\left(\mu \nn + \alpha \jj \right)\Nspin.  
\end{equation}
We revert here to the earlier notation $\nn$ for $\nn(\usebox{\horline})$;
${\nn}(\usebox{\urline})$ and ${\nn}(\usebox{\ulline})$ can be recovered
from $\nn$ and $\jj$ as
\begin{equation}
  \nonumber
  2\bar{\nn}(\usebox{\urline})
  = {{\nn+\jj}},
  \quad
   2\bar{\nn}(\usebox{\ulline})
  = {{\nn-\jj}}.
%  = {\textstyle\frac{n-j}{2}}
\end{equation}
%%%%%%%%%%%
%  $|j| \le n$
%%%%%%%%%%%
We recover the entropy as a function of the densities via Legendre transform.
That is, using our established notation
\hbox{$\phi(\mu,\alpha) = \Nspin^{-1}\ln\partfn(0,\mu,\alpha)$} for the
free entropy,
\begin{equation}
  \label{eq:n-and-j}
\nn = \frac{\partial \phi}{\partial \mu},
\qquad
\jj = \frac{\partial \phi}{\partial \alpha},  
\end{equation}
for $1\ll C,L$ the entropy per spin $s(\nn,\jj)$ is 
\begin{equation}
  \label{eq:Legendre-s-phi}
  s(\nn,\jj)
=  \phi(\mu,\alpha)  - \mu {\nn} - \alpha \jj.
\end{equation}
Note that $\nn$ and $\jj$ are in a sans-serif font to avoid any
confusion with generic summation variables later.

The partition function (\ref{eq:Z-mu-alpha}) can be considered
the Lagrangian formulation of the model. The usual Lagrangian
coupling of an electromagnetic vector potential to current density
is $\int i\boldsymbol{A}\cdot\boldsymbol{j}$.
Our $i\alpha$ should be thought of as an imaginary vector potential.

\subsubsection{Translation to fermion language.}

Tranlation of the partition function (\ref{eq:Z-mu-alpha}) 
into fermionic language requires a minor variation on
what was done in section \ref{sec:xfer_0}.
To give the correct weights to the movements of particles,
we generalize the condition (\ref{eq:T0-conditions-x})
imposed on the transfer operator $\xfer_0$ to 
\begin{equation}
\xfer_\alpha c_x^\dagger = (e^{-\alpha}c_{x-1}^\dagger + e^{\alpha}c_{x}^\dagger) \xfer_\alpha.
\label{eq:T-alpha}
\end{equation}
Proceed just as in section \ref{sec:xfer_0}. First,
\begin{equation}
\xfer_\alpha\, c(q)^\dagger  = 2\cos \left({\textstyle \frac{q}{2}} +i\alpha \right) 
e^{iq/2} {c}(q)^\dagger \xfer_\alpha,
\label{eq:T-alpha-c-dagger}
\end{equation}
gives $\alpha$-dependent pseudo-energies
$\varepsilon_\alpha(k)$ defined by [cf. (\ref{eq:mode-energy})] 
\begin{equation}
  \nonumber
\varepsilon_\alpha(k)
=  \varepsilon(k+2\alpha i)
= -\ln\left[ 2\cos \left({\textstyle\frac{k}{2}} + i \alpha \right) \right].
\end{equation}
In contrast to the Lagrangian form in (\ref{eq:Z-mu-alpha}), we now have a Hamiltonian
form, and we see that $i\alpha$ shows up in the expected form for a vector potential.

The logarithm of the partition function is
\begin{equation}
{\ln Z_{_{C}}({L}^{-1},\mu,\alpha)}
= CL \phi_{_{C,L}}(\mu,\alpha) 
= \sum_{|q|}  \ln \left[ 1 + e^{(\mu -\varepsilon_\alpha(q)) L}\right].
\end{equation}
In the limit $L\to\infty$, terms with $\mu < \mathrm{Re}\, \varepsilon_\alpha(q)$
tend to zero, and therefore, in the full thermodynamic limit $C,L\to\infty$,
the density is implicitly given by
\begin{equation}
\label{eq:mu}
\mathrm{Re}\, \varepsilon_\alpha(n\pi) = \mu,
\end{equation}
and the free entropy density of the spin system by
\begin{equation}
\label{eq:n-alpha-free-entropy}
\phi(\mu,\alpha)
= \mu n - \int_{0}^{n\pi} \mathrm{Re}\, \varepsilon_\alpha(q) \, \frac{dq}{\pi}.
\end{equation}
We wish to obtain $s(\nn,\jj)$ from this by Legendre transformation (\ref{eq:Legendre-s-phi}).
Eqs. (\ref{eq:n-and-j}) give the transformation from control parameters
$(\mu,\alpha)$ to densities $(\nn,\jj)$ abstractly.
The relation (\ref{eq:mu}) begins to make it concrete. To finish the job, note that,
from (\ref{eq:n-and-j}), 
\begin{eqnarray}
\nonumber 
\jj & = \frac{\partial\phi}{\partial \alpha}
=
\pi \frac{\partial n}{\partial\alpha}\left(\mu - \mathrm{Re}\, \varepsilon_\alpha(n\pi)\right)
- \int_{-n\pi}^{n\pi} \frac{\partial \varepsilon_\alpha(q)}{\partial \alpha} \, \frac{dq}{2\pi}
\nonumber \\
& = - \int_{-n\pi}^{n\pi}  2i \frac{\partial \varepsilon_\alpha(q)}{\partial q} \, \frac{dq}{\pi}
               = {\frac{2}{\pi}} \mathrm{Im}\, \varepsilon_\alpha(n\pi).
\end{eqnarray}
Putting things together, we obtain a simple, pretty, and useful relation:
\begin{equation}
  \label{eq:alpha+mu/n+j}
  2 \cosh \left( \alpha - i {\textstyle\frac{\pi}{2}} n \right)
  = \exp \left( - \mu - i {\textstyle\frac{\pi}{2}} {j}\right).
\end{equation}
Some manipulation produces the disentangled form
\numparts
\begin{eqnarray}
  &  \tan {\textstyle\frac{\pi \jj}{2}}
    = \tanh \alpha \cdot \tan {\textstyle\frac{\pi \nn}{2}},
    \label{eq:nj-mualpha-1}
  \\
& e^{-2\mu}  = 4(\cos^2 {\textstyle\frac{\pi \nn}{2}}  + \sinh^2 \alpha). 
    \label{eq:nj-mualpha-2}
\end{eqnarray}
\endnumparts
Summing up, the entropy density is
\begin{equation}
  \label{eq:s(n,j)-integral}
  s(\nn,\jj) = -\alpha \jj
  + \mathrm{Re}\, \int_{i2\alpha}^{\nn\pi+i2\alpha}
  \ln\left(2\cos {\frac{z}{2}}\right) \, \frac{dz}{\pi}, 
\end{equation}
where (\ref{eq:nj-mualpha-1}) determines $\alpha$ in terms of $(\nn,\jj)$,
while $\mu$ has simply dropped out.
The following section will detail ways to evaluate the integral here.
Results are plotted in Fig.~\ref{fig:entropy-surface}, and it is indeed
strictly concave.
Each slice through the graph of Fig.~\ref{fig:entropy-surface}
from a corner to the centre of the opposite edge is identical
to Fig.~\ref{fig:entropy-curve}.

%%%%%%%%%%%%%%%%%%%%%%%%
\subsection{Evaluation of integrals}\label{sec:integrals}

This section gives some details on ways to evaluate the integral in
(\ref{eq:s(n,j)-integral}), of which (\ref{eq:s(n)-integral}) is a special case.
The results are Eqs. (\ref{eq:s(n,j)}) and (\ref{eq:s(n)-with-Bernoulli}).

  Simply rearranging things, we have
\begin{equation}
  \pi[ s(\nn,\jj) +\alpha\jj ]
  =\Delta(\alpha,\nn)
  \defeq \mathrm{Re}\,[I(-i2\alpha+\pi \nn) - I(-i2\alpha)],
\end{equation}
where
\begin{equation}
  I(z) \defeq \int^z \ln\left(2\cos\frac{w}{2}\right)\, dw.
\end{equation}
To avoid having to worry about the zeros of cosine on the
real axis, assume our contours are all in either the upper or lower half-plane.
Now, because we need to preserve only the real part of the difference for two
integration terminal points with the same imaginary part, we are free to transform
this integral in a variety of ways, such as
(i) move the initial point, (ii) add an imaginary constant or
imaginary multiple of $w$ to the integrand.
Therefore, since
$\cos{\textstyle\frac{w}{2}}
= \sin{\textstyle\frac{\pi-w}{2}} = ie^{-i(\pi-w)}(1-e^{i(\pi-w)})$,
\begin{equation}
I(z) \circeq - \int^{\pi - z} \ln\left(1-e^{iw}\right)\, dw.  
\end{equation}
Here, we use $\circeq$ to mean that the real part of the difference between
the two sides depends only on the imaginary part of the argument.
In other words, they are equivalent for our purposes.
Notice that as we assumed $z$ in the lower half-plane, our contours are now
in the upper half-plane.
We consider two series representations of this integral.

\subsubsection{``Large'' $|\alpha|$.}

Since $|e^{iw}| \le 1$ for $\mathrm{Im}\, w \ge 0$, we may take our reference
point at $+i\infty$ and apply the series expansion for logarithm to obtain
\begin{eqnarray}
I(z) & \circeq -\int^{\pi-z} \ln\left(1-e^{iw}\right)\, dw  
        = \sum_{m=1}^\infty \frac{1}{m} \int^{\pi-z} e^{imw} \, dw
        \nonumber \\
& = i \sum_{m=1}^\infty \frac{e^{im (\pi-z)}}{m^2}.
\end{eqnarray}
Now, this last expression is purely imaginary 
whenever $\mathrm{Re}\, z$ is a multiple of $\pi$, so
\begin{eqnarray}
  \label{eq:s(n,j)}
& s(\nn,\jj) =
- \alpha \jj
+  \sum_{k=1}^\infty \frac{1}{k^2} e^{-2|\alpha|k}\sin k\pi \bar{\nn},
\\
& \mathrm{where}\quad
     \alpha =  \tanh^{-1} \frac{\tan {\textstyle{\pi\jj}/{2}}}{\tan {\textstyle{\pi \nn}/{2}}}
     \nonumber
\end{eqnarray}
The series converges for all $\alpha$,
more quickly the larger $|\alpha|$, yet it is reasonably practical even for $\alpha=0$.

%%%%%%%%%%
\subsubsection{Small $|z|$.}
Now we consider an alternative expansion around zero.
It is genuinely limited, with a radius of convergence of $2\pi$,
but converges very rapidly for small $z$.
Integrate by parts as
\begin{equation}
\ln\left(1-e^{iw}\right) =  
\frac{d}{dw}\left[ w\ln\left(1-e^{iw}\right) \right]   
- \frac{-iw}{e^{-iw}-1},
\end{equation}
and recognize the second term here as the generating function
of Bernoulli numbers:
\begin{equation}
  \frac{-iw}{e^{-iw}-1} = 1 -i\frac{w}{2} - \sum_{m=1}^\infty {|B_{2m}|}\frac{w^{2m}}{(2m)!}.
\end{equation}
We recall that $B_0 = 1$, $B_1 = 1/2$, and
\begin{equation}
\sum_{k=0}^{m-1} {m \choose k} B_k = 0, \;\mathrm{for}\; m>1.
\end{equation}
Then, we have, for $|u| < 2\pi$,
\begin{equation}
  I(\pi-u) \circeq -u\ln\left({2}\sin{\textstyle\frac{u}{2}}\right)
  + u  - \sum_{k=1}^\infty \frac{|B_{2k}|}{(2k+1)!}  {u^{2k+1}}.
\end{equation}
This is especially handy for $\alpha=0$ (equivalently, $\jj=0$), in which case
\begin{equation}
  \label{eq:s(n)-with-Bernoulli}
  s(\nn) = -\bar{\nn}\ln\left({\frac{2}{e}}\sin{\frac{\pi\bar\nn}{2}}\right)
 - \sum_{k=1}^\infty \frac{|B_{2k}|\pi^{2k}}{(2k+1)!}  {\bar{\nn}^{2k+1}}.
\end{equation}
Here, we have fixed the constant by recognizing that $s(0) = s(1) = 0$,
as is clear from the physical meaning.

To bound the tail of the series, use (23.1.15) of
Abramowitz and Stegun\cite{Abramowitz+Stegun}:
\begin{equation}
|B_{2m}| < 4 \frac{(2m)!}{(2\pi)^{2m}}  
\end{equation}
This implies that the series in (\ref{eq:s(n)-with-Bernoulli})
is dominated term-by-term, by this one:
\begin{equation}
4\ln\frac{1+\bar{\nn}}{1-\bar{\nn}} = 8 \sum_{m=1}^\infty \frac{\bar{\nn}^{2m+1}}{2m+1}. 
\end{equation}
The bounds reported in Section~\ref{sec:entropy-series} came from this.

%%%%%%%%%%%%%%%%%%%%%%%%%%%%%%%%%
\section{Nonzero temperature}\label{sec:T>0}

This section extends the fermionic representation of the TAFIM to nonzero temperature.
Keep in mind that the temperature of the fermion system is not
the temperature of the spin system; $T$ always means the latter.
The temperature of the fermion system is the
reciprocal of the length of the lattice on which the spin system is defined.
As we are primarily interested in the thermodynamic limit, this means that the
fermion system continues to be at zero temperature here. Instead, the temperature
$T$ corresponds to a change of the fermion Hamiltonian or time-evolution operator.
Our translation is not exact.
In the interest of a fermionic description which is elementarily tractable,
we make an approximation which captures the leading-order (roughly speaking)
behavior for nonzero temperature, as (\ref{eq:thermo-asymptotics}) demonstrates
conclusively.
Nevertheless, Fig.~\ref{fig:approx-vs-exact-thermo} shows that
it turns out to be a relatively good approximation even up to high
temperature.
Our approximation will have the form of a reduced BCS model for
superconductivity. The approximation consists in neglecting more
complicated interactions, rather than from a mean field approximation
as in real BCS theory. Thus, there is no transition to a normal state
on raising the fermion temperature, which is equivalent to shortening
the cylinder on which the model lives.

%%%%%%%%%%%%%%%%%%% 
\begin{figure}[h]
  \centering
  \includegraphics[width=85mm]{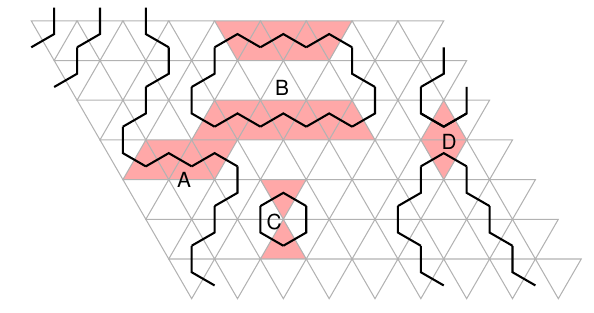}
  \caption{A relative string diagram for nonzero temperature.
    ``Excited'' triangles, containing three unsatisfied bonds, are highlighted.
    The approximation analyzed here disallows neighboring excited triangles, as at
    \textsf{A} and \textsf{B}, but pairs meeting on a horizontal bond, as in \textsf{D},
  are permitted. Thus, the approximation does not preserve complete rotational symmetry.}
  \label{fig:T>0-diagram}
\end{figure}
%%%%%%%%%%%%%%%%%%%%%%%
The critical difference of $T > 0$ is that the motif \usebox{\creation}
has a nonzero weight, hence string diagrams can contain \textit{closed loops},
as in Fig.~\ref{fig:T>0-diagram}.
A priori, the motif \usebox{\creation} comes with a weight
$\kappa^2 \defeq e^{-4\beta J}$.
With boundary conditions enforcing the same number of particles at initial
and final times, we can make things a little more symmetrical
by exploiting the fact that we then have an equal number of
\usebox{\creation} and \usebox{\annihilation} motifs.
One can verify this by inspection of Fig.~\ref{fig:T>0-diagram}.
Then, assigning weight $\kappa$ to both $\usebox{\creation}$
and $\usebox{\annihilation}$ gives the same weight for all
string diagrams. This is just for convenience, however, and we
could do without it.

The small loop at \textsf{C} in Fig.~\ref{fig:T>0-diagram}
corresponds to creation of a pair (\usebox{\creation}) followed by its
immediate annihilation (\usebox{\annihilation}).
At \textsf{D}, there is
an annihilation event followed by a creation event one time step later.
Unfortunately, completely general identification of \usebox{\creation} with pair-creation
and \usebox{\annihilation} with pair-annihilation is not possible, as we see from
the loop labeled \textsf{B}, which is one event of each kind, and even more so from
\textsf{A}, which has neither creation nor annihilation.
Our basic approximation is to neglect these processes and keep only isolated
\usebox{\creation} and \usebox{\annihilation} motifs, in which case the
interpretation is sound --- with one caveat, explained in Section~\ref{sec:approximate-T}.

From the statistical mechanical perspective, \textsf{A} and \textsf{B}
are suppressed not just energetically, but also entropically; excited
triangles should not clump together.
From the Hamiltonian perspective of the fermion formulation, 
they correspond to small corrections to the transfer matrix, being higher
order in $\kappa$, and they represent complicated many-body interactions
which are not qualitatively new processes.
To see why these are many-body interactions, note that
a well-separated pair can annihilate, as at \textsf{B}, only
if the intervening region is free of particles.
It is plausible that our approximation captures the leading
$T>0$ behavior, in some sense. This is verified in a precise
way in (\ref{eq:thermo-asymptotics}).

\subsection{Approximate transfer matrix}\label{sec:approximate-T}

In previous sections, $\xfer_0$ implemented a time step $\Delta \tau = 1$ by
moving, or not moving, each particle: $\Delta x = -1,0$.
Now, there are two additional categories of option. The full transfer
matrix $\xfer$ should first implement the annihilation option, then the motion
option, then the creation option, in that order, because if a pair is annihilated
or created, the particles composing it do not have the option of moving in that
time step.
That is, the transfer matrix is 
\begin{equation}
\label{eq:xfer-3-factor}
\xfer = \xfer_{\mathrm{pr}}^\dag \xfer_0 \xfer_{\mathrm{pr}},
\end{equation}
where $\xfer_{\mathrm{pr}}$ and $\xfer_{\mathrm{pr}}^\dag$ implement pair destruction
and creation, respectively. With abbreviation
\begin{equation}
\kappa \defeq e^{-2\beta J},  
\end{equation}
they are given by
\begin{equation}
\label{eq:Tpr}
\xfer_{\mathrm{pr}}
= \prod_{i\in\Latt} (1 + \kappa c_{i+1} c_i) \quad \mathrm{and} \quad
% \end{equation}
% \begin{equation}
% \label{eq:pair-creation-0}
\xfer_{\mathrm{pr}}^\dag
= \prod_{i\in\Latt} (1 + \kappa c_{i}^\dag c_{i+1}^\dag).
\end{equation}

As with $\xfer_0$, these are put into a more useful form in $k$-space.
Since the operators $\setof{c_{i+1}c_i}{i \in\Latt}$
are nilpotent, $1+\kappa c_{i+1} c_i = \exp(\kappa c_{i+1} c_i)$;
and since they commute with each other
$\xfer_{\mathrm{pr}} = \prod_{i\in\Latt} e^{\kappa c_{i+1} c_i}$.
Applying commutativity again produces
$\xfer_{\mathrm{pr}} = e^{-H_{\mathrm{pr}}}$, where
\begin{equation}
\label{eq:T-pair-destruction}
H_{\mathrm{pr}}  
= - \kappa \sum_{i \in\Latt}  c_i c_{i+1}
= \kappa \sum_{0 < q\in\BZ} 2i\sin q \cdot b(q),
\end{equation}
where
\begin{equation}
  \label{eq:b(q)}
b(q) \defeq c(-q) c(q)
\end{equation}
destroys a pair of fermions in the $q$ and $-q$ modes.
Re-expanding, we finally obtain
\begin{equation}
\label{eq:Tpr(q)}
\xfer_{\mathrm{pr}}  = \prod_{0<q\in\BZ}[ 1 + 2i \kappa \sin q\cdot  b(q)].
% \end{equation}
% The pair-creation part of $\xfer$ can now be written down immediately:
% \begin{equation}
%   \label{eq:xfer-pr-dag}
%% superfluous?
% \quad
% \xfer_{\mathrm{pr}}^\dag  = \prod_{0<q\in\BZ}[ 1 - 2i \kappa \sin q\cdot  b(q)^\dag].
\end{equation}

\begin{figure}[h]
  \centering
  \includegraphics[width=80mm]{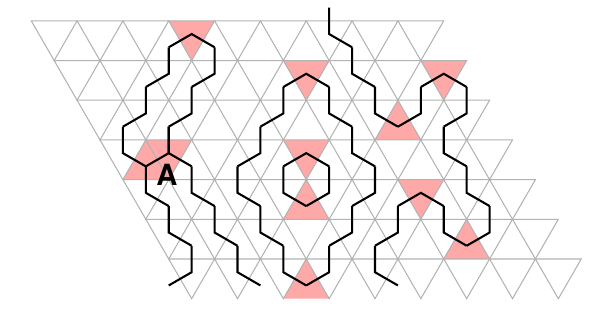}
  \caption{A string diagram generated by the approximate transfer matrix
    (\ref{eq:xfer-3-factor}). Shaded triangles indicate factors of $\kappa$.
    In addition to the intended processes, $\xfer$ also generates spurious
    events of the sort labeled {\textsf {\bfseries A}}, with a weight $\kappa^2$.
    Here, a \usebox{\annihilation} and \usebox{\creation} motif
    are juxtaposed in a way which is not legitimate in the original model.
  }\label{fig:T>0-oops}
\end{figure}

Now that we have presented the mathematical forms, it is easier to see that,
not only have we neglected some processes, but have also included some
undesired ones.
Fig.~\ref{fig:T>0-oops} illustrates a string diagram generated by the transfer
matrix (\ref{eq:xfer-3-factor}). Unfortunately, it generates some unwanted events,
such as at {\textsf{\textbf A}}. It represents a spurious correlated pair-hopping
process with a strength $\kappa^2$.
Though unwelcome, it plausibly does not have a large effect on the results.
%%%%%%%%%%%%%%%%%%% 
%%%%%%%%%%%%%%%%%%%%%%%%%%%%%%%

\subsection{BCS-style fermion pairing}\label{sec:BCS}

The transfer matrix (\ref{eq:xfer-3-factor}) can be written as a product
\begin{equation}
\xfer = \prod_{0 \le q \in\BZ} {\xfer}(q)
\label{eq:T-factorized}
\end{equation}
of commuting factors, where $\xfer(q)$ is the part affecting the
$q$ and $-q$ fermion modes. Explicitly, for $q > 0$,
\begin{eqnarray}
\xfer(q) & = [ 1 - 2i \kappa \sin q\cdot  b^\dag] e^{-\varepsilon n}[ 1 + 2i \kappa \sin q\cdot  b]
  \nonumber \\
& = [ 1 - 2i \kappa \sin q\cdot  b^\dag] [e^{-\varepsilon n} + 2i \kappa \sin q\cdot  b]
  \nonumber \\
 & =  e^{-\varepsilon n}
 + 4\kappa^2 \sin^2 q\cdot  b^\dag b
 + 2i \kappa \sin q\cdot  (b-b^\dag).
\label{eq:xfer(q)}
\end{eqnarray}
Here, we have suppressed the $q$ labels and $n = c(q)^\dag c(q) + c(-q)^\dag c(-q)$
counts the number of particles in those two modes.

$\xfer(q)$ is the sum of a term $\xfer_{\mathrm{o}}(q)$ (`o' for `odd')
acting on the subspace with one particle in either $q$ or $-q$,
which is simply the constant $e^{-\varepsilon}$,
and a term $\xfer_{\mathrm{e}}(q)$ acting on the even-particle-number subspace.
$\xfer_{\mathrm{e}}(q)$ is conveniently represented in terms of 
Pauli matrices $\tau_\alpha$.
With ``spin-up'' corresponding to both modes $q$ and $-q$ occupied,
\begin{eqnarray}
\xfer_{\mathrm{e}}(q)
% = {A} + B\Big[\tau_z \cos\theta + \tau_y \sin\theta \Big]
& = \left(  \begin{array}{cc}
    e^{-2\varepsilon}+4\kappa^2\sin^2 q   &  -2i\kappa\sin q
    \\
    -2i\kappa\sin q & 1
  \end{array} \right)
                      \nonumber \\
&  = e^{i\frac{\theta}{2}\tau_x}\Big(
{A} + B\tau_z 
\Big) e^{-i\frac{\theta}{2}\tau_x}.
\label{eq:q-q-2-particle}
\end{eqnarray}
Here (restoring the argument $q$),
\begin{equation}
\label{eq:A(k)}
{A}(q)
= {\textstyle\frac{1}{2}} \Tr \xfer_{\mathrm{e}}(q)
% = {\textstyle \frac{1}{2}} \left(e^{-2\varepsilon}+ 4\kappa^2 \sin^2 q + 1\right)
% \nonumber \\ & \;\;
= {\textstyle\frac{3}{2}} + \cos q + 2\kappa^2 \sin^2 q,
\end{equation}
while $B(q)$ is the positive square root of
  \begin{equation}
\label{eq:B(k)}
  B(q)^2 = \det \xfer_e(q) - A(q)^2 = (A(q) - 1)^2 + 4\kappa^2\sin^2 q
\end{equation}
  and $\theta(q) \in [0,\pi]$ is determined by
  \begin{equation}
 \cos\theta(q) =  {\textstyle\frac{A(q)-1}{B(q)}}.
\end{equation}
Note for future reference that $A(k_F) = 1$, where $k_F = \frac{2\pi}{3}$ is
the Fermi point. It is evident that, at least for small $\kappa$, $\cos\theta(k)$
goes monotonically from $1$ at $k=0 $ to $-1$ at $k=\pi$, crossing zero near $k_F$.
%%%%%%%%%%%%%%%%%%% 

The \textit{larger} eigenvalue of $\xfer_{\mathrm{e}}(q)$ is $A+B$, with corresponding
eigenvector
\begin{equation}
  \label{eq:u-v}
\left( \begin{array}{c} v_q \\ u_q \end{array} \right)
=  e^{ i\frac{\theta}{2}\tau_x }
\left( \begin{array}{c} 1 \\ 0 \end{array} \right)
  =
  \left( \begin{array}{c}
%           \cos{\textstyle\frac{\theta}{2}} \\
%           i\sin{\textstyle\frac{\theta}{2}}
           \cos{\theta/2} \\
           i\sin{\theta/2}
  \end{array} \right)
\end{equation}
Thus, the ground state takes the familiar BCS form
\begin{equation}
  \label{eq:BCS-state}
  \ket{\Omega(\kappa)}
  = \prod_{0 < k\in \mathrm{BZ}}
  \left( u_k  + v_k b(k)^\dag \right)
\ket{\Vac}.
\end{equation}
with energy density
\begin{equation}
  \label{eq:superconducting-ground-energy}
  \frac{E}{C}
  = -\ln\braket{\Omega}{\xfer}{\Omega}
  = -\frac{1}{C}\sum_{q>0} \ln [{A}(q) + B(q)].
\end{equation}
We have ignored the $k=0$ mode, and will continue to do so.
It occurs only for odd $\NN$ anyway, and does not matter in the thermodynamic limit.

%%%%%%%%%%%%%%%%
\subsection{Thermodynamic functions}
%%%%%%%%%%%%%%%
\begin{figure}
  \centering
  \includegraphics[width=80mm]{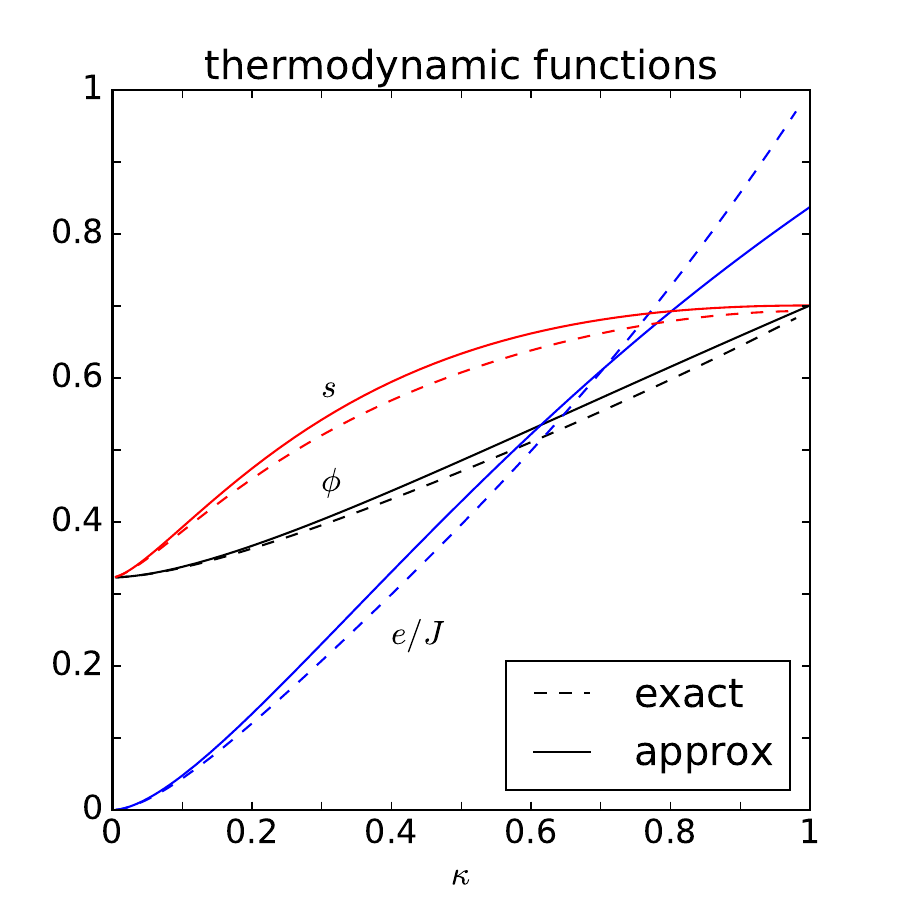}
  \caption{Thermodynamic potentials per spin.
    $\phi$ is the Helmholtz free entropy per spin $\Nspin^{-1}\ln \partfn$,
    $e$ is energy density in excess of $T=0$ value, $s$ is the entropy density, and
    $\kappa = e^{-2\beta J}$.
    Solid lines indicate exact values\cite{Wannier-50,Houtappel-50,Lavis+Bell-v1},
    and dashed lines, the approximation developed in this Section.
    Note that
    $\frac{e}{J}-1$ is three times the difference between densities of
    satisfied and unsatisfied bonds, making clear why $e/J\to 1$ as $T\to\infty$.
  }
  \label{fig:approx-vs-exact-thermo}
\end{figure}
%%%%%%%%%%%%%%%%%%%

Armed with the ground state (\ref{eq:BCS-state}) and its energy
(\ref{eq:superconducting-ground-energy}), we compute the thermodynamic functions,
energy density $e/J$, entropy density $s$ and free entropy density $\phi$, related by
\begin{equation}
  \label{eq:free-entropy-s-e}
\phi = s - \beta e.  
\end{equation}
We calculate below $\phi$ and $e$ in our approximate theory,
and obtain $s$ through the above.
The results are plotted in Fig.~\ref{fig:approx-vs-exact-thermo},
along with exact results.
For example, from Lavis \& Bell\cite{Lavis+Bell-v1}, 8.167 (8.13.4 in the first edition)
the exact energy density (with ground state value subtracted) is
%\begin{eqnarray}
\begin{equation}
  \label{eq:e/J-Lavis+Bell}
  \frac{e}{J}
=  1 - \frac{1+\kappa^2}{1-\kappa^2}
%    \nonumber \\
  + \frac{1}{\pi}\frac{d\eta}{d\kappa}\int_{-1}^{1}  \frac{dx}{\sqrt{(1-x^2)[(x+\eta)^2-2x-2] }}
  \end{equation}
%\end{eqnarray}
where
\begin{equation}
\eta = \frac{1}{2}\, \frac{3+\kappa^4}{1-\kappa^2}.  
\end{equation}

%%%%%%%%%%%%%%
\subsubsection{Free entropy}

Just as at zero temperature, the limiting free entropy per spin,
\begin{equation}
\phi 
 = \lim_{C,L\to\infty} \Nspin^{-1}\ln\partfn(T)
 = \frac{1}{2}\int_0^\pi  \ln [{A}(q) + B(q)]   \, \frac{dq}{\pi}.
  \label{eq:free-entropy-T>0}
\end{equation}
is equal to minus the ground state energy density of the corresponding fermion system,
which is given in (\ref{eq:superconducting-ground-energy}).
At both extremes, $\kappa = 0,1$, the energetic contribution to $\phi$
vanishes --- at zero temperature because all states have the same energy,
and at infinite temperature because $\beta\to 0$ while $e$ is bounded.

%%%%%%%%%%%%%%
\subsubsection{Energy}

Basic statistical mechanics says that the energy per spin, $e$, satisfies
\begin{equation}
  \label{energy=phi-derivative}
  \beta e
  = -\beta\frac{\partial\phi}{\partial\beta}
  = 2\beta J \kappa \frac{\partial\phi}{\partial\kappa}.
\end{equation}
This derivative can be calculated directly from the formula in
(\ref{eq:free-entropy-T>0}). We will take a more interesting route.
Since $C\phi$ equals the ground state energy $E$ of the fermion system,
and $\braket{\Omega}{\xfer}{\Omega} = e^{-E}$, the Hellmann-Feynman theorem
gives
\begin{equation}
  \label{eq:Hellmann-Feynman}
  \frac{\partial}{\partial\kappa} C\phi
 = \frac{\partial}{\partial\kappa}\ln\braket{\Omega}{\xfer}{\Omega}
  = \frac{\braket{\Omega}{\partial\xfer/\partial\kappa}{\Omega}}%
    {\braket{\Omega}{\xfer}{\Omega}}
\end{equation}
Now, from (\ref{eq:Tpr(q)}), and using that
the various $b(q)$ square to zero and commute with each other,
\begin{equation}
  \frac{\partial}{\partial\kappa}\xfer_{\mathrm{pr}}
=  \xfer_{\mathrm{pr}}\sum_n c_nc_{n+1},
\end{equation}
and similarly for $\xfer_{\mathrm{pr}}^\dag$. Since, moreover,
$\Omega$ is a translation-invariant eigenstate of $\xfer$,
(\ref{eq:Hellmann-Feynman}) is rewritten as
\begin{equation}
  \label{eq:Hellmann-Feynman-2}
\frac{\partial\phi}{\partial\kappa} 
= -{2}\braket{\Omega}{c_n c_{n+1}}{\Omega}
\end{equation}
Therefore,
\begin{eqnarray}
  \frac{e}{J}
  \label{eq:e/J}
  &= 2\kappa \braket{\Omega}{c_n c_{n+1}}{\Omega}
  %  \nonumber
  \\
  &= 4 i \kappa \int_0^\pi \sin q\, \braket{\Omega}{c(q) c(-q)}{\Omega}\, \frac{dq}{\pi}
    \nonumber \\
  &= \BZave{ 2 \kappa \sin q\, \sin\theta(q) }.
    \nonumber 
\end{eqnarray}
Evidently, the energy density is closely related with the superconducting order parameter.

%%%%%%%%
\subsubsection{Low-temperature asymptotics}

The asymptotic behavior of the thermodynamic functions in the vicinity
of the critical point $T=0$ is of especial interest.
We work out the leading behavior here, starting with the energy, as it 
has a particularly convenient form.
We cannot say that
\begin{equation}
\sin\theta(k) = \frac{2\kappa\sin k}{B(k)}
\end{equation}
is simply ${\mathcal O}(\kappa)$ because the denominator has 
a zero at $(\kappa=0,\, k=k_F)$.
The Fermi point is $k_F = \frac{2\pi}{3}$, so that
$\frac{1}{2} + \cos k_F = 0$, and the expansion of $B(k_F + q,\kappa)$
in (\ref{eq:B(k)}) to ${\mathcal O}(q^2,\kappa^2)$ yields
\begin{equation}
  \nonumber
B(k_F+q)^2 = (\sin^2k_F)  (q^2 + 4\kappa^2) + \cdots,
\end{equation}
and then
\begin{equation}
\sin\theta(k_F+q) \sim \frac{2\kappa}{\sqrt{q^2+4\kappa^2}}.
\end{equation}
On the other hand,
$B(k)$ is nonzero even at $\kappa=0$, away from $k_F$, so there
are no other singularities.
Therefore,
\begin{equation}
  \nonumber
\frac{e}{J} 
\sim 2\kappa \frac{\sqrt{3}}{2}
\int_{-\epsilon}^\epsilon \frac{2\kappa}{\sqrt{q^2+4\kappa^2}}\,\frac{dq}{\pi}
\sim  \frac{4\sqrt{3}}{\pi} \kappa^2 \ln \frac{\epsilon}{\kappa}.
\end{equation}
Here, $\epsilon$ is some cutoff which does not really matter since we will
throw away the $\ln\epsilon$ term.
Using (\ref{energy=phi-derivative}) and \hbox{$\phi=s-\beta e$},
we then deduce the following set of asymptotics:
\begin{eqnarray}
  \label{eq:thermo-asymptotics}
\frac{e(\kappa)}{J} &  \sim  \frac{4\sqrt{3}}{\pi} \kappa^2 |\ln {\kappa}|
% \nonumber
  \\
{\phi(\kappa)} &  \sim  s(0) + \frac{\sqrt{3}}{\pi} \kappa^2 |\ln {\kappa}|
                \nonumber \\
{s(\kappa)} &  \sim  s(0) + \frac{3\sqrt{3}}{\pi} \kappa^2 |\ln {\kappa}|^2.
                \nonumber 
\end{eqnarray}
The residual zero-temperature entropy density is given in (\ref{eq:residual-entropy}).
The above asymptotic formulas are in precise agreement with exact results.
To see this, check that the integrand in (\ref{eq:e/J-Lavis+Bell}) has a singularity at
\hbox{$\kappa=0$}, \hbox{$x=-1/2$}
of the same sort as the one analyzed above, and the coefficient is the same.
Since $e$ and $s(0)$ determine everything else, the asyptotics in (\ref{eq:thermo-asymptotics}) are exact.
A natural approach to a low-temperature expansion directly on the two-dimensional
model would treat triangles with three unsatisfied bonds as elementary excitations.
Since each such triangle has a Boltzmann weight $\kappa^2 = e^{-4\beta J}$,
we might have anticipated that the first terms would be simply proportional
to $\kappa^2$. Since that sort of reasoning works fine for systems with ordered
ground states, the appearance of logarithms in (\ref{eq:thermo-asymptotics})
is a surprise from this perspective.
We have no intuitive justification for logarithms as such, but presumably they
must be ascribed to the fact that the background into which the excitations
are inserted has a nontrivial structure.

%%%%%%%%%%%%%%%%%%%%%%%%%%%%%%%%%%%%%%%%5
\section{Spin correlation functions}\label{sec:spin-spin-correlator}

\subsection{Correlation lengths and energy gaps}
\label{sec:correlation-lengths-energy-gaps}

Suppose $A(x,\tau)$ is TAFIM observable which can be expressed as a local
operator $\hat{A}$ in the fermion interpretation. An example is bond energy
for circumferential bonds, which is 
$J(1-2n_x)$, with $n_x$ the number of particles at site $x$.
For observables of this sort, we have
\begin{eqnarray}
  \langle A(x,\tau) A(y,\tau+s)\rangle
& =
  \frac{\braket{\Omega}{\hat{A}(x)e^{-sH}\hat{A}(y)}{\Omega}}
    {\braket{\Omega}{e^{-sH}}{\Omega}}
 \\
&=\sum_{m}
   \braket{\Omega}{\hat{A}(x)}{\Phi_m}
    \braket{\Phi_m}{\hat{A}(y)}{\Omega}e^{-(E_m-E_0) s},
    \nonumber %\\
\end{eqnarray}
where $\Phi_0 \equiv \Omega$ is the ground state.
Thus, the connected correlation function is
\begin{eqnarray}
\expct{ A(x,\tau) A(y,\tau+s) } & - \expct{A(x,\tau)}\expct{A(y,\tau+s)}
    \nonumber \\
& =\sum_{m > 0}
    \braket{\Omega}{\hat{A}(x)}{\Phi_m}
    \braket{\Phi_m}{\hat{A}(y)}{\Omega}e^{-(E_m-E_0) s}.
    \nonumber 
\end{eqnarray}
The rate at which this correlation function falls off
is the fundamental energy gap $E_1-E_0$ above the ground state, or
even faster if the relevant matrix elements vanish.
The energy gap is zero when ($\kappa = 0$) the fermion system is a free fermi gas.
When $\kappa > 0$, the gap is the energy to create a quasiparticle,
and if $\kappa$ is very small, the lowest-energy quasiparticles have
momentum close to $k_F$. 
From (\ref{eq:A(k)}) and (\ref{eq:B(k)}),
$A(k_F) = 1 + {\textstyle\frac{3}{2}}\kappa^2$,
and
$ B(k_F) = \sqrt{3}\kappa$
Therefore, by
(\ref{eq:superconducting-ground-energy}),
\begin{equation}
{E_1-E_0} \sim \sqrt{3}\kappa.
\end{equation}
This corresponds to a correlation length diverging as $e^{2\beta J}$,
showing that $T=0$ is a critical point.
Some connected correlations might decay at a faster rate, but none
will be slower.

%%%%%%%%%%%%%%% 
  \subsection{Spin correlations in terms of satisfied bonds}
  \label{sec:spin-correlations-satistied-bonds}

  Stephenson\cite{Stephenson-64,Stephenson-70a} found that the
  spin-spin correlation function in the \textit{strict} TAFIM does indeed behave
  in the way suggested by the preceding analysis:
  Along a bond direction $\hat{e}$, asymptotically as $\ell\to\infty$,
%  \begin{equation}
%  \langle \sigma_i\, \sigma_{i + \ell \hat{e}}\rangle
%    \sim e^{-2 \kappa \ell} \cos{\textstyle\frac{2\pi\ell}{3}}
%  \label{eq:spin-spin-T>0-Stephenson}
%  \end{equation}
%   for $T>0$
%  \begin{equation}
%  \langle \sigma_i\, \sigma_{i + \ell \hat{e}}\rangle
%  \sim \ell^{-1/2} \cos n\pi\ell,
%  \label{eq:spin-spin-Stephenson}
%  \end{equation}
%  %
%   for $T>0$ along a bond direction $\hat{e}$, and
\begin{eqnarray}
\langle \sigma_i\, \sigma_{i + \ell \hat{e}}\rangle
&  \sim e^{-2 \kappa \ell} \cos{\textstyle\frac{2\pi\ell}{3}}, \qquad T > 0
  \label{eq:spin-spin-T>0-Stephenson}
\\                                                      
\langle \sigma_i\, \sigma_{i + \ell \hat{e}}\rangle
& \sim \ell^{-1/2} \cos n\pi\ell, \qquad T = 0
\label{eq:spin-spin-Stephenson}
\end{eqnarray}
However, the spins do not correspond to local fermionic operators,
so it is not immediately clear what this has to do with the analysis
of Section \ref{sec:correlation-lengths-energy-gaps}.
Indeed, the very notion of a spin-spin correlation function seems to make
little sense from our general perspective.
For instance, changing the sign of all bonds attached to $\sigma$ leaves the
system fully frustrated, but changes the sign of $\langle \sigma \sigma' \rangle$
for every other spin $\sigma'$.
This is why we emphasized that Stephenson's results were for the \textit{strict}
TAFIM, with every bond AFM.
We want to make sense of the spin correlation function from the point of
view of general fully frustrated systems not merely for the sake of
generality, but also because the computational strategy is based on
that understanding.

For the strict TAFIM (planar or not), we can rephrase the product $\sigma\sigma'$
of spins as $\Gamma(\Cyc{C}) \defeq (-1)^{| \Cyc{C}^{\Sat} |}$,
where $\Cyc{C}$ is any path of bonds from $\sigma$ to $\sigma'$.
Hence, we can rephrase the spin correlation function in terms of satisfied bonds,
a characterization which fits our point of view, and which can be generalized.
In general, if $\Cyc{C}$ and $\Cyc{C}'$ are two paths between $\sigma$ and $\sigma'$,
then $\Cyc{C}+\Cyc{C}'$ is a loop, and we have
\begin{equation}
  \nonumber
  |\Cyc{C}^{\Sat}| + |\Cyc{C}'^{\Sat}|
  \eqpar
  |(\Cyc{C}+\Cyc{C}')^{\Sat}|
  \eqpar 
|(\Cyc{C}+\Cyc{C}')_{\FM}|.
\end{equation}
Therefore, the rewrite of the spin product as $\Gamma(\Cyc{C})$
generalizes painlessly to all fully frustrated bond patterns for which
FM is trivial in cohomology. 
On its face, $\Gamma(\Cyc{C})$ is very nonlocal, since it is expressed in
terms of an entire path, but since it really only depends on the end points
of $\Cyc{C}$, that is to some extent illusory.

But, how are we to think of $\Gamma(\Cyc{C})$ and
its expectation value $\langle \Gamma(\Cyc{C}) \rangle$?
If $|\Cyc{C}^{\Sat}|$ had the same value in every configuration,
then $|\langle \Gamma(\Cyc{C}) \rangle|$ would be one.
So, smaller values of that expectation would seem to indicate greater
fluctuations in $\Cyc{C}^{\Sat}$. In fact, however, a lot of
information is lost in passing to $\Gamma(\Cyc{C})$ since it only depends
on the parity of $|\Cyc{C}^{\Sat}|$.
We really should free ourselves from the presumption that only the parity
matters. Studying fluctuations of $|\Cyc{C}^{\Sat}|$ for various paths
$\Cyc{C}$ is a very natural way of studying, possibly even conceiving of,
configurational fluctuations. Our approach will follow this philosophy.
For $\Cyc{C}$ a path along a lattice direction of variable length $\ell$,
we obtain first suitable information on the distribution of 
$|\Cyc{C}^{\Sat}|$ itself (it's gaussian for large $\ell$), and then
extract $\langle \Gamma(\Cyc{C}) \rangle$ from that information.

In the thermodynamic limit, the correlation function along all
three bond orientations is the same.
The fermionic representation is especially suited to calculating
$\Gamma(\Cyc{C})$ straight along what it regards as the spatial direction.
In that case, $|\Cyc{C}|$ is the number of particles along the path.
The spin correlation function therefore
is related to fluctuations in particle density.
Following that approach,
we reproduce Stephenson's result (\ref{eq:spin-spin-Stephenson})
for zero temperature,
while extending it to particle densities different from $2\pi/3$:
\begin{equation}
\langle \sigma_i\, \sigma_{i + \ell \hat{e}}\rangle
\sim \ell^{-1/2} \cos n\pi\ell.
\label{eq:spin-spin}
\end{equation}
At $T > 0$, we obtain Stephenson's functional form, but with
a slightly different coefficient in the exponent:
\begin{equation}
\langle \sigma_i\, \sigma_{i + \ell \hat{e}}\rangle
  \sim e^{-\frac{\pi^2}{4} \kappa \ell} \cos{\textstyle\frac{2\pi\ell}{3}}.
\label{eq:spin-spin-T>0}
\end{equation}
We are unsure what to make of the discrepancy.

%%%%%%%%%%%%%%%%

\subsection{Approach to spin-spin correlations}
\label{sec:spin-correlator-idea}

Now we begin to fill in details of the calculation of
(\ref{eq:spin-spin}) and (\ref{eq:spin-spin-T>0}.
Taking $\hat{e}$ along the circumferential direction,
and remembering that a vertical string segment represents a satisfied bond,
we have
\begin{equation}
\sigma_i \sigma_{i+\ell{\hat{e}}} = e^{i\pi N_\ell},
\nonumber
\end{equation}
where
\begin{equation}
N_\ell \defeq \sum_{j=1}^{\ell} n_j
\end{equation}
is the number of particles in the segment $[1,\ell]$ at some fixed time.
Then,
\begin{equation}
  \label{eq:spin-spin-to-numbers}
  \langle \sigma_i \sigma_{i+\ell{\hat{e}}} \rangle
%= \left\langle \exp\left\{i\pi\sum_{j=0}^{\ell-1} n_j\right\}\right\rangle,
= \left\langle e^{i\pi N_\ell} \right\rangle
\defeq \bra{\FS} e^{i\pi N_\ell} \ket{\FS},
\end{equation}
where $\Omega$ is the fermionic ground state.
Splitting $N_\ell$ into a mean and fluctuation as  
\begin{equation}
  \nonumber
N_\ell  = {\ell}n + \Nfluct, 
\end{equation}
we obtain
\begin{equation}
  \nonumber
\left\langle e^{i\pi N_\ell} \right\rangle 
=  e^{i\pi n\ell} \expct{e^{i\pi \Nfluct}}.
\end{equation}
Since $N_\ell$ is integral, the left-hand side is necessarily real,
hence so is the right hand side. However, if the exact value of
$\expct{e^{i\pi \Nfluct}}$ is replaced by an asymptotic expression, that property may
be lost, so we will be safe and take the real part explicitly:
\begin{equation}
\left\langle e^{i\pi N_\ell} \right\rangle 
=  \mathrm{Re}\, e^{i\pi n\ell} \expct{e^{i\pi \Nfluct}}.
\end{equation}
Now, assuming the fluctuations ($\Nfluct$) are asymptotically gaussian, 
$\langle e^{i\pi \Nfluct} \rangle \sim \exp[- \frac{\pi^2}{2} \langle \Nfluct^2 \rangle]$,
and therefore
\begin{equation}
\label{eq:characteristic-fn}
\left\langle e^{i\pi N_\ell} \right\rangle 
\sim
(\cos n\ell)\, e^{ - \frac{\pi^2}{2} \langle \Nfluct^2 \rangle }.
\end{equation}

The program therefore is to compute $\langle \Nfluct^2 \rangle$,
show that $\Nfluct$ has a normal distribution, and thence recover 
(\ref{eq:spin-spin-T>0}) and (\ref{eq:spin-spin}).

%%%%%%%%%%%%%%%%%%%%%
\subsubsection{Poisson summation}\label{sec:Poisson}

The preceding is evidently based on the fact that
if $X$ is a random variable with gaussian probability density function
\begin{equation}
  \nonumber
\gamma(x) = \frac{e^{-(x-\mu)^2/2\sigma^2}}{\sqrt{2\pi\sigma^2}},
\end{equation}
then
\begin{equation}
  \expct{e^{i\xi X}} = \hat{\gamma}(\xi) \defeq \int e^{-i\xi x} \gamma(x)\, dx
  = e^{i\xi\mu -\frac{\sigma^2}{2} \xi^2}.
\end{equation}
the Fourier transform of $\gamma$.
However, if $N$ takes only integer values, the distribution of
$N$ cannot really be gaussian, and that presumably explains how
$\expct{e^{-i 3 \pi N}}$ can equal $\expct{e^{-i\pi N}}$.

The resolution of the conundrum is to say that $N$ takes value
$m$ with probability proportional to $\gamma(m)$.
Then,
\begin{equation}
  \nonumber
  \expct{e^{-i\theta N}} \propto \sum_{m\in \Int} e^{-i\theta m} \gamma(m),
\end{equation}
and by the Poisson summation formula, the right-hand sum is equal to
\begin{equation}
  \label{eq:Poisson}
\sum_{m\in\Int} \hat{\gamma}(\theta + 2\pi m).
\end{equation}
This solves the problem because for large $\sigma^2$, the sum in
(\ref{eq:Poisson}) is dominated by the term (or two in exceptional cases)
with minimal $|\theta + 2\pi m|$. In short, for $\expct{e^{i\theta N}}$,
we should always reduce $\theta$ to the interval $(-\pi,\pi]$.

%%%%%%%%%%%%%%
\subsubsection{Number variance}

To determine the variance of $\Nfluct$, proceed as follows.
Convert the number operators to momentum representation:
\begin{eqnarray}
  n_x &= \frac{1}{C}\sum_{x} e^{ikx} \rho(k),
        \nonumber \\
\rho(k) &= \sum_{x} e^{-ikx} n_x.
  \nonumber
\end{eqnarray}
Rewrite the number fluctuation operator as
\begin{equation}
\Nfluct = \frac{1}{C}\sum_{k \in \BZO} w(k) \rho(k), 
\end{equation}
with
\begin{equation}
w(k) =  e^{ik(\ell+1)/2}\, \frac{\sin {k\ell}/{2}}{\sin {k}/{2}}.
\end{equation}
Then,
\begin{equation}
\label{eq:N-var-sum}
\langle \Nfluct^2 \rangle
= \frac{1}{C^2}\sum_{q \in \BZO} |w(q)|^2 \|\rho(q)\Omega\|^2.
\end{equation}

We proceed to evaluate this successively for $T=0$, and 
then for $T > 0$.

%%%%%%%%%%%%%%%%
\subsection{Zero temperature}\label{sec:correlation-T=0}

This subsection shows that, at zero temperature,
\begin{equation}
\label{eq:N-variance}
\mathrm{Var}\; N_\ell 
= \expct{ \Nfluct^2 }
\sim \frac{1}{\pi^2}{\ln\ell}.
\end{equation}
Substituted into (\ref{eq:characteristic-fn}), this yields exactly
the claimed result (\ref{eq:spin-spin}).
However, the supposition that the fluctuations are normal (i.e., gaussian),
is cast into some doubt by the $\ell$ dependence of $\expct{ \Nfluct^2 }$.
A normal distribution is usually explained as the result of many (nearly) independent
small contributions.
For an ordinary, say classical, particle system we expect fluctuations
of particle number in mesoscopic subvolumes to be those small contributions,
leading to the variance of $N$ being proportional to system size, as in
ordinary thermodynamic fluctuation theory.
Evidently, the one-dimensional fermion system at zero temperature does not behave that way.
Subsection \ref{sec:normal} shows that, nevertheless $\Nfluct$ does have a normal
distribution.

%%%%%%%%%%%%%%%%%% 
\subsubsection{Number variance.}\label{sec:N-variance}

This subsection is devoted to a careful and detailed derivation of (\ref{eq:N-variance}).
The treatment has similarities to that of Villain and Bak\cite{Villain81}.

Acting on the Fermi sea $\FS$,
$\rho(q)$ creates particle-hole excitations of momentum $-q$.
If $|q|$ is small, there are $|q|C/2\pi$ distinct
(that is, orthogonal) such excited states.
Also,
\hbox{$\langle \rho(-q^\prime)\rho(q)\rangle = \langle \rho(q^\prime) \FS | \rho(q) \FS \rangle$}
is zero unless $q=q^\prime$. Hence, as long as $q < 2k_F, 2(\pi-k_F)$, 
\begin{equation}
  \expct{ \rho(-q) \rho(q) }
= \| \rho(q) \Omega \|^2
= \frac{C|q|}{2\pi},
\label{eq:rho-rho-expectation}
\end{equation}
%   \begin{equation}
%   q < 2k_F, 2(\pi-k_F) 
%   \;\Rightarrow\;
%    \langle \rho(-q) \rho(q) \rangle  = \frac{C|q|}{2\pi},
%   \label{eq:rho-rho-expectation}
%   \end{equation}
and therefore, from (\ref{eq:N-var-sum},
%\begin{eqnarray}
\begin{equation}
\label{eq:N-var-sum-T=0}
  \langle \Nfluct^2 \rangle
%  &= \frac{1}{C^2}\sum_{q \in \BZO} |w(q)|^2 \|\rho(q)\Omega\|^2
%    \nonumber \\
  \lesssim\frac{1}{C}\sum_{q \in \BZO} \frac{|q|}{2\pi} |w(q)|^2.
\end{equation}
%\end{eqnarray}
The inequality arises because the contribution of large $|q|$ is overestimated.
However, we will see that the important contribution arises from small $|q|$, so
in the thermodynamic limit
\begin{equation}
  \label{eq:N-var-integral}
\langle \Nfluct^2 \rangle
\simeq
\frac{1}{2\pi^2} \int_{0}^\pi q |w(q)|^2 {dq}.
%= \frac{1}{2\pi^2}I_0^{{\pi}/{2}}(\ell).
\end{equation}
In order to clearly isolate the singularity
it is useful to introduce the family of integrals
\begin{equation}
\label{eq:I-integral-defined}
 I_{\alpha}^{\beta}(\ell)
\defeq \int_{2\alpha}^{2\beta} |w(k)|^2 k\, dk
= 4 \int_{\alpha}^{\beta} \sin^2 {\ell x} \, \frac{x\, dx}{\sin^2 {x}}
\end{equation}
for \hbox{$0\le \alpha < \beta \le \frac{\pi}{2}$}.
We will show that, for $\ell\to\infty$, 
\begin{equation}
  \label{eq:I-asymptotics}
  I_\alpha^\beta(\ell) =
  \left\{
    \begin{array}{ll}
    2 \ln \ell + {\Cal O}(1) & \alpha = 0,
    \\
{\Cal O}(1) & \alpha > 0,
  \end{array}\right.
\end{equation}
to obtain formula (\ref{eq:N-variance}) for the number variance
$\expct{\Nfluct^2}$.

In (\ref{eq:I-asymptotics}), the case $\alpha > 0$ is clear, so we concentrate
on $\alpha = 0$.
With \hbox{$\theta: [0,\pi/2] \rightarrow [0, 1-(2/\pi)^2]$} a monotone continuous function,
substitute
\begin{equation}
  \nonumber
  \frac{1}{\sin^2 x}
  = x^{-2}[1 + x^2\theta(x)]
  = \frac{1}{x^{2}} + \theta(x)
\end{equation}
into the integral to obtain
\begin{equation}
\nonumber
I_0^\beta(\ell) 
= 4 \int_0^{\ell\beta} \frac{\sin^2 y}{y} \, dy
+ 4 \int_0^{\beta} x \theta(x) \sin^2 {\ell x}\, dx.
\end{equation}
The second integral above is plainly ${\Cal O}(1)$.
For the first, split the integration range and apply the
sine double-angle formula to reduce it to
\begin{equation}
  \nonumber
\int_0^{\beta} \frac{\sin^2 y}{y} \, dy   
     + \frac{1}{2}\int_{\beta}^{\ell\beta} \frac{dy}{y}
 - \frac{1}{2}\int_{\beta}^{\ell\beta} \cos 2y \frac{dy}{y}.
\end{equation}
The second integral is the important one --- it evaluates to $\ln \sqrt{\ell}$.
The first is a well-defined constant (independent of $\ell$), and the
third is also ${\Cal O}(1)$, as can be seen after integration by parts:
\begin{equation}
\int_{\beta}^{\ell\beta} \cos 2y \, \frac{dy}{y}
=   \frac{\sin 2y}{2y} \Big|_{\beta}^{\beta\ell} 
+ \frac{1}{2} \int_{\beta}^{\beta\ell} \frac{\sin 2y}{y^2}\, dy.
\nonumber
\end{equation}
This completes the demonstration of (\ref{eq:I-asymptotics}).

%%%%%%%%%%%%%%%% 
\subsubsection{Normality.}\label{sec:normal}

Now, we turn to the verification that $\Nfluct$ is due to many weakly dependent modes so
that it really has a Gaussian distribution.
To get better control of $\langle e^{i\theta\Nfluct}\rangle$, 
we split the number fluctuation into parts 
$\tilde{N}_{>}$ and $\tilde{N}_{<}$ 
containing the density modes with wavevectors in $(0,\pi)$ and $(-\pi,0)$, respectively.
Since $\Nfluctp^\dagger = \Nfluctm$, and recalling that all the density modes commute, we have
\begin{equation}
\left\langle \FS \,\Big|\, e^{i\theta \tilde{N}} \FS \right\rangle 
= 
\left\langle e^{-i\theta \Nfluctp} \FS \,\Big|\, e^{i\theta \Nfluctp} \FS \right\rangle.
\label{eq:expectation-as-inner-prod}
\end{equation}
Proceed to examine the state $e^{i\theta\Nfluctp} \FS$. 
The part, ${\Cal H}$, of the fermion model Hilbert space which is relevant to us 
has a precise particle number (the size of the Fermi sea). 
This Hilbert space can be decomposed as
${\Cal H} = \bigoplus_{n=0}^\infty {\Cal H}_n$,
according to the number of particle-hole excitations, 
$n$ (number of particles missing from the Fermi sea).
Now, $\Nfluctp \,\FS$ is in ${\Cal H}_1$, of course.
Applying a second fluctuation operator, we see that
$\Nfluctp^2 \,\FS \in {\Cal H}_2 \oplus {\Cal H}_1$;
the second $\Nfluct^+$ may create a second particle-hole pair, or it may move the
already excited particle.
The crucial point to note, though, is that 
the norm of the component in ${\Cal H}_1$ is ${\Cal O}(C^{-1})$ times that of
the component in ${\Cal H}_2$.
This is because $\rho(q)$ can create $qC/2\pi$ distinct particle-hole
excitations, but there is only one way it can boost the already excited
particle. The argument just given can be applied over and over, 
showing that for any $m \ll C$, $(\Nfluctp)^m\,\FS$ is in ${\Cal H}_m$,
to relative order $C^{-1}$. 
Therefore, the component of $(\Nfluctp)^m\,\FS$
orthogonal to ${\Cal H}_m$ is negligible in the thermodynamic limit.

It will now be useful to split a Fourier component $\rho(q)$ of the density 
as $\rho(q) = \rho(q)^+ + \rho(q)^- + \rho(q)^0$, where $\rho(q)^+$ 
(the particle-hole creation part) contains the terms moving a particle from 
inside the Fermi sea to outside, $\rho(q)^- = [\rho(-q)^+]^\dagger$, those moving 
a particle from outside to inside, and $\rho(q)^0$ is the remainder. 
Correspondingly, we write
\begin{equation}
\Nfluctp^+ = \frac{1}{C}\sum_{q>0} w(q) \rho(q)^+,
\end{equation}
and so forth. With this new notation, we can write the conclusion reached
in the previous paragraph as
\begin{equation}
(\Nfluctp)^m  \,\FS \simeq (\Nfluctp^+)^m  \,\FS \in {\Cal H}_m.
\end{equation}
Here is below, `$\simeq$' means `up to relative ${\Cal O}(C^{-1})$'.
Returning now to the inner product in (\ref{eq:expectation-as-inner-prod})
that we need to compute,
\begin{equation}
\left\langle e^{-i\theta \Nfluct^+} \FS \,\Big|\, e^{i\theta \Nfluct^+} \FS \right\rangle
\simeq \sum_m \frac{(-\theta^2)^m}{(m!)^2} \| (\Nfluctp^+)^m \,\FS \|^2.
\label{eq:double-exponential-expansion}
\end{equation}
It only remains to evaluate the norm squared, $\| (\Nfluctp^+)^m \,\FS \|^2$,
occuring on the right, which is immediately rewritten as
\begin{equation}
  \nonumber
\langle (\Nfluctp^+)^{m-1} \,\FS \,|\, \Nfluctm^- (\Nfluctp^+)^{m} \,\FS \rangle.
\end{equation}
To evaluate this, the commutator $[\Nfluctm^-,\Nfluctp^+]$ is needed so
that $\Nfluctm^-$ can be moved through until it annihilates $\Omega$.
The key to this is
\begin{eqnarray}
[\rho(q)^-,\rho(k)^+] \simeq \frac{C c(q)}{2\pi}\delta(q,-k),
\nonumber \\
\mathrm{where} \quad c(q) = |q| \;\;\mbox{if}\;\, |q| < 2k_F, 2\pi-2k_F,  
\end{eqnarray}
by essentially the same reasoning leading to Eq. (\ref{eq:rho-rho-expectation})
for $\langle \rho(q)\rho(k)\rangle$.
Insofar as the sought-for ``many weakly dependent'' modes are made explicit here,
this is the linchpin of the calculation.
Adding up over wavevector yields
\begin{equation}
  \nonumber
[\Nfluctm^-,\Nfluctp^+] \sim \frac{I_0^{\pi/2}(\ell)}{4\pi^2},
\end{equation}
and therefore 
\begin{equation}
\Nfluctm^- (\Nfluctp^+)^{m} \,\FS 
\simeq m  \frac{I(\pi,\ell)}{4\pi^2}(\Nfluctp^+)^{m-1} \,\FS.
\nonumber
\end{equation}
With (\ref{eq:I-asymptotics}) a simple recursion now gives 
\begin{equation}
\| (\Nfluctp^+)^m \,\FS \|^2 \sim  m!  \left(\frac{\ln \ell}{2\pi^2}\right)^m. 
\end{equation}
Substituting into (\ref{eq:double-exponential-expansion}) produces the 
power series of an exponential. Provisionally, we arrive at
\begin{equation}
\label{eq:provisional}
\expct{e^{i\theta\tilde{N}}}
\sim \exp \left( - \frac{\theta^2}{\pi^2} \ln \sqrt{\ell} \right). 
\end{equation}
If this holds for all $\theta\in\Real$, it indeed verifies that
$\Nfluct$, and therefore $N_\ell$ have gaussian distributions.
Substitution of $\pi$ for $\theta$ yields the claimed result
(\ref{eq:characteristic-fn}).

%%%%%%%%%%%%%%%%%%%%%%
\subsection{Correlation length at nonzero temperature
}\label{sec:correlation-T>0}

The method to determine the correlation length for $T>0$ is essentially the same,
namely an examination of density fluctuations, and application of
(\ref{eq:N-var-sum}), using the
approximate superconductor theory to determine asymptotics of
$\expct{ \Nfluct^2 }$ for small $\kappa > 0$, as $\ell\to\infty$.
The result of the following calculation is
\begin{equation}
  \label{eq:N-var-SC}
\expct{\Nfluct^2} \sim \frac{\kappa}{2}\ell.
\end{equation}
Comparison with (\ref{eq:characteristic-fn}) implies an exponential decay
$\expct{\sigma_{n}\sigma_{n+\ell}} \sim e^{-\ell/\xi}$, with
inverse correlation length
\begin{equation}
\xi^{-1} \sim \frac{\pi^2}{4} \kappa.
\end{equation}

To express the density operators in a useful way, we need another
basic tool of superconductor theory, namely, the Bogoliubov-Valatin transformation
\begin{equation}
\label{eq:Bogolibov-Valatin}
\gamma(k) = u_kc(k)-v_kc(-k)^\dag
\end{equation}
expressing the quasiparticles operator in terms of \hbox{$c$-fermions}.
The inverse of the preceding is
\begin{equation}
  \label{eq:BV-inverse}
c(k) = \bar{u}_k\gamma(k)-v_k\gamma(-k)^\dag,
\end{equation}
where we take
\begin{equation}
  \label{eq:symmetry-u-v}
u_{-k} = -u_k, \quad v_{-k}=v_k,  
\end{equation}
corresponding to $\theta(k)$ being odd in $k$ [See (\ref{eq:u-v})].

Rewrite $\rho(q)=\sum_{k\in\mathrm{BZ}} c(k+q)^\dag c(k)$ using the Bogoliubov-Valatin
transformation and notice that $\|\rho(q)\Omega\|^2$ only has contributions
from the pieces with two quasiparticle creation operators:
\begin{equation}
%  \rho(q)  &=\sum_{k\in\text{BZ}} c(k+q)^\dag c(k)
%  \nonumber \\
%  &=\sum_{k\in\text{BZ}}
c(k+q)^\dag c(k) = -u_{k+q}{v}_k \gamma(k+q)^\dag \gamma(k)^\dag + \cdots.
\end{equation}
Thus,
\begin{equation}
\nonumber %  \label{eq:rho(q)Omega}
\|\rho(q)\Omega\|^2 = \sum_{k\in\BZ} |u_{k+q}{v}_k|^2.
\end{equation}

Compared to the zero-temperature ($\kappa = 0$) situation considered in
Section~\ref{sec:spin-spin-correlator}, an important difference of 
$\kappa > 0$ is that the above expression has a regular expansion at $q=0$
with a nonzero constant term. To capture the leading part of
$\expct{\Nfluct^2}$ with respect to $\ell$, we therefore simply set $q=0$.
With $|u_{k}{v}_k|^2 = {\textstyle\frac{1}{4}} \sin^2\theta(k)$, 
\begin{equation}
  \label{eq:rho(q)Omega-2}
  \|\rho(q)\Omega\|^2
  \simeq \frac{C}{4} \int \sin^2\theta(k)\, \frac{dk}{\pi},
\end{equation}
independent of $q$, as long as it is small. 

\begin{equation}
  \int_0^\pi \sin^2\theta(k) dk
  \sim \int_{-\infty}^\infty \frac{4\kappa^2}{q^2+4\kappa^2}\, dq
  = 2\pi \kappa.
\end{equation}

Turning to the other component of (\ref{eq:N-var-sum}),
\begin{eqnarray}
\int_{\mathrm{BZ}} |w(q)|^2 \frac{dq}{2\pi}  
    &= \int_0^\pi \frac{\sin^2(q\ell/2)}{\sin^2(q/2)} \frac{dq}{\pi}
      \nonumber \\
    &\sim \frac{2\ell}{\pi}\int_0^\infty \frac{\sin^2y}{y^2} {dy}
%      \nonumber \\
    = \ell.
\end{eqnarray}
Assembling the pieces yields
\begin{equation}
  \nonumber
\expct{\Nfluct^2} \sim \frac{1}{4\pi} 2\pi\kappa \ell = \frac{\kappa}{2}\ell,
\end{equation}
as announced in (\ref{eq:N-var-SC}).

%%%%%%%%%%%%%%%%%%%%%%%%%%%55
\section{Conclusion}

The TAFIM is a basic model of statistical mechanics and the prototype of
frustration, and for that reason is invoked by physicists in a wide variety of contexts.
We have given a unified treatment of many aspects of TAFIM physics, relying
on a simple mapping to a system of fermions in one space, one (imaginary)
time dimension.
This permits easy exact calculation and elucidation of many properties
at zero temperature, which is anyway where the TAFIM is special.
For example, this strategy easily uncovers the strange two-dimensional
continuum of equilibrium macrostates and their entropy densities.
A novel perspective on the divergence of the spin correlation length
at zero temperature also emerges. From the fermionic perspective, this
is where a superconducting gap closes and number fluctuations pass from
extensive to logarithmic in system size.

%%%%%%%%%%%%%%%%%%%%%%%

%%%%%%%%%%%%%%%%% 
%  \ack
%  This project was funded by the National Science Foundation under
%  awards DMR-1420620 and DMR-2011839.
%  AN acknowledges funding from the University of Akron.
%  
%%%%%%%%%%%%%

\section*{References}
% \bibliographystyle{iopart-num}
% \bibliography{frustchannel}
%%%%%%%%%
\providecommand{\newblock}{}

%%%%%%%%%

%%%%%%%%%%%%5
\end{document}